\documentstyle[preprint,aps,epsf]{revtex}
\begin{document}

\title{The Weak Decay of Hypernuclei}

\author{A. Parre\~no, A.\ Ramos}
\address{Departament d'Estructura i Constituents de la Mat\`eria,
Facultat de F\'{\i}sica,\\
Diagonal  647, E-08028 Barcelona, Spain}

\author{C.\ Bennhold}
\address{Center of Nuclear Studies, Department of Physics,\\ The George
         Washington University, Washington, D. C. 20052}

\date{\today}
\maketitle

\begin{abstract}

The nonmesonic weak decay of $\Lambda$ hypernuclei is studied in a
shell model framework. A complete strangeness-changing weak $\Lambda N \to NN$
transition potential, based on one boson exchange,
is constructed by including the exchange of
the pseudoscalar mesons $\pi$, K, $\eta$ as well as
the vector mesons $\rho, \omega$, and K$^*$, whose weak coupling
constants are obtained from soft meson theorems and SU(6)$_w$.
General expressions for nucleons in arbitrary shells are obtained.
The transition matrix elements include realistic
$\Lambda$N short-range correlations and NN final state interactions
based on the Nijmegen baryon-baryon potential.
The decay rates are found to be especially sensitive to the inclusion of
the strange mesons, K and K$^*$, even though the
role of kaon exchange is found to be reduced with recent couplings
obtained from next-to-leading order Chiral Perturbation Theory.
With the weak couplings used
in this study the rates remain dominated by the pion-exchange mechanism
since the contributions of heavier mesons either cancel each other or
are suppressed by form factors and short-range correlations.
The total decay rate therefore remains in agreement with present
measurements.  However, the partial rates which are even more sensitive
to the inclusion of heavier mesons cannot be reconciled with the
data.  The proton asymmetry changes by 50\% once
heavier mesons are included and agrees with the available data.
\end{abstract}

\section{Introduction}
In single $\Lambda$ hypernuclei, a $\Lambda$ hyperon can occupy any
orbital in the hypernucleus since it is free from the Pauli exclusion
principle due to its additional quantum number Strangeness.
Hypernuclei are typically produced in some excited state through
hadronic reactions such $(K^-,\pi^-)$ or $(\pi^+, K^+)$ but can reach
their ground state through electromagnetic gamma and/or nucleon
emission.
Eventually, they will decay through weak interaction processes which
involve the emission of pions or nucleons but are nonleptonic in nature.

A free $\Lambda$-hyperon has a lifetime of
about 260 picoseconds and decays almost totally into a
pion and a nucleon ($\Lambda \to p \pi^-$ ($\sim 64$\%),
$\Lambda \to n \pi^0$ ($\sim 36$\%)), with a release of kinetic energy
of about 5 MeV to the nucleon along with a corresponding
final momentum of about 100 MeV/c.
When the $\Lambda$ is embedded in the nuclear medium,
the phase space for the mesonic decay is greatly reduced since the
$\Lambda$ is bound by some 10 MeV for p-shell hypernuclei and up to 30
MeV in heavy hypernuclei. The final-state nucleon with its very low
momentum thus becomes Pauli blocked, leading to a suppression of
the mesonic rate by several orders
of magnitude for heavy hypernuclei such as
$^{208}_\Lambda$Pb. Experimentally, however, one finds the
lifetimes of
hypernuclei to be roughly independent of $A$ (see, e.g., Fig. 14-1 in
Ref. \cite{bando}),
though the data base is very poor, especially for systems with $A > $12.
Therefore, the nuclear medium surrounding the bound $\Lambda$ affects
its weak decay by introducing new, nonmesonic decay modes, such as $\Lambda N
 \to NN$.
Thus, hypernuclei larger than $^{5}_\Lambda$He decay mainly through
these nonmesonic channels, where the $\Lambda$ mass excess of 176 MeV
is converted into kinetic energy of a final state of nucleons
emerging with a momentum of about 400 MeV/c.
As a result, this process has a much larger phase space relative to the
mesonic one, and the outgoing nucleons are not Pauli blocked.

This was recognized more than forty years ago\cite{primakoff} when it was
suggested that the novel two-baryon decay mode $\Lambda N \to NN$ could be
 understood
in terms of the free-space decay mechanism $\Lambda \rightarrow \pi N$ with
the exception that the pion now has to be considered virtual and is absorbed
on a second nucleon bound in the hypernucleus. However, the large momentum
transfer involved in the reaction leads to a mechanism that is sensitive to
the short distance behavior of the amplitude and thus raises the possibility
that the exchange of heavier mesons may play an important role. The
production of these mesons would be below threshold for the free-space
$\Lambda$ decay, but they can contribute through virtual exchange in a
two-baryon decay channel.  As discussed further below
it is the kinematic freedom of these additional
boson exchanges that provides part of the motivation for this study.
The fairly large momentum transfer also raises the hope that this reaction
turns out to be
insensitive to nuclear structure details and thus creates a suitable
channel to investigate the weak decay mechanism.

We would like to point out that there is
another possible nonmesonic decay channel,
the two-nucleon induced process
$\Lambda N N\to NNN$, where the virtual pion emitted at the weak vertex is
absorbed by a pair of nucleons which are correlated through the strong
force.  This mechanism was first
investigated in Ref.~\cite{albe} where it was suggested that its
magnitude could be comparable to $\Lambda N \to NN$.  However, a reanalysis with
 more
realistic assumptions\cite{ramos1,ramos3} reduced its contribution to 10-15\%
of the total nonmesonic decay rate. Its relevance lies mainly in its
potential to renormalize the experimentally measured partial ratios;
we will return to this point later when discussing results.

The main goal in studying the weak nonmesonic decay channel is to gain
insight into the fundamental aspects of the four-fermion, strangeness
changing weak interaction.  Most of the earlier work on
nonleptonic weak processes proceeded directly from some model weak
Hamiltonian and computed experimental observables that could then be
compared to measurements. More recently the approach has been divided
into a two-stage process.  The first step starts with the Standard Model
electroweak Hamiltonian at the sub-hadronic level and takes into account
QCD corrections at short distances, yielding a so-called effective weak
Hamiltonian. In relation to the mass of the W-boson of 80 GeV, these
would be low momentum transfer processes leading to a zero-range or
contact interaction. The Cabbibo theory combined with strong-interaction
corrections would result in $V-A$ weak interaction and presumably
predict the relative strength of the $\Delta S$=0 and $\Delta S$=1
transition. Thus, hadronic weak matrix elements of the form
$\langle M B' | H_w \mid B  \rangle$ could eventually be calculated.
The second stage
involves using these weak vertices as a starting point for effective
nuclear two-body operators that are then implanted into the nucleus
with the usual nuclear many-body wave functions. It is the latter task
which is the main subject of this paper.

The most important information regarding the nonleptonic weak
interactions of hyperons comes from their free-space decays. However, while
these processes have been well-measured and understood phenomenologically,
a more basic understanding in terms of the underlying degrees of freedom is
still lacking.  The situation is similar for the nonmesonic decay modes,
except that the experimental data are more unsatisfactory. The early
measurements were based on bubble chamber experiments and emulsion works,
who suffered from low precision, poor statistics and difficulties with the
identification of the particular hypernucleus, except for the very light
hyperfragments.  These experiments were nevertheless able to establish the
first limits on hypernuclear lifetimes, albeit with large uncertainties.
Their techniques were mostly aimed at measuring the emission of the $\pi^-$
from the mesonic decay channel; the analyses of the nonmesonic decay modes
were greatly inhibited by the low hypernuclear production rates, the limited
spacial resolution of early detectors and the presence of one or more neutral
particles (especially neutrons) in the final state. In recent years a series
of counter experiments carried out at BNL and KEK improved the quality of
data on the nonmesonic decay modes using pion and kaon beams. In contrast
to these measurements LEAR at CERN explored hypernuclei in the A=200 mass
region following antiproton annihilation\cite{Lear}.
The heavy hypernuclei were
identified through their delayed fission events which were attributed to the
nonmesonic decays.  While this interpretation is fraud with difficulties
it resulted in lifetimes similar to the those of
p-shell hypernuclei, thus raising the
intruiging possibility that the $\Lambda N \to NN$ mechanism saturates already
 in the A=12
system.  On the other hand, employing the ($K^-,\pi^-$) production
mechanism with direct timing techniqes, the Brookhaven AGS measurements
resulted in better data on the hypernuclear lifetimes and branching ratios
of mesonic as well as nonmesonic decay channels of $^{12}_\Lambda$C,
 $^{11}_\Lambda$B and the
different helium hypernuclei\cite{szymanski,schuma}. The group at KEK
used the
($\pi^+,K^+$) production reaction to not only measure total and partial decay
rates\cite{noumi}, but also to use the induced polarization in $^{12}_\Lambda$C
 to
determine
for the first time the difference in the number of protons emitted
along the axis of polarization compared to the number ejected in the opposite
direction\cite{ajim}. This
asymmetry is a direct consequence of the presence of parity-violating
components in the weak decay mechanism.

Parity violation in hadronic systems represents a unique tool
to study aspects of the nonleptonic weak interaction between hadrons.
The nonmesonic process
resembles the weak $\Delta$S=0 nucleon-nucleon interaction that has been
explored experimentally in parity-violating NN scattering
measurements
by measuring the asymmetry of longitudinally polarized protons.
However, the $\Lambda N \to NN$ two-body decay mode contains more information
 since
it can explore both the parity-conserving (PC) and the parity-violating
(PV) sector of the $\Delta$S=1 weak baryon-baryon interaction while in
the weak NN system the strong force masks the signal of the weak
PC interaction.  On the other hand, the free process $\Lambda N \to NN$ cannot
 be
accessed experimentally which complicates the interpretation of the
nonmesonic decay rates since the reaction mechanism has to be studied in
the environment of hypernuclear structure.

A number of theoretical approaches to the $\Lambda N \to NN$ decay mode have
 been developed
over the last thirty years which are more extensively reviewed in
Ref.~\cite{cohen1}. The early phenomenological analyses
by Dalitz et. al.\cite{dalitz} provided the general nonrelativistic structure
of the $\Lambda N \to NN$ amplitude which was then related to decay rates of
 s-shell
hypernuclei using certain simplifying assumptions.  The $\Delta S=0$ weak
nucleon-nucleon interaction at low and intermediate energies has generally
been described in a meson exchange model involving one strong interaction
vertex and one weak one; the same basic assumption has been used for a
microscopic description of the $\Delta S$=1 $\Lambda N \to NN$ mechanism. Early
 calculations
based on the one-pion exchange (OPE)
were due to Adams\cite{adams}, modifications
of the OPE due to strong interactions in the nuclear medium were suggested in
Ref.~\cite{salcedo} to account for many-body nuclear structure effects.
At the very least, the OPE mechanism can be expected to adequately
descibe the long-range part of the $\Lambda N \to NN$ interaction. The
first attempts to include heavier bosons, at first the $\rho$ meson,
--- again in complete analogy to the $\Delta S$=0 NN interaction ---
were presented in Ref.~\cite{mckellar,nardulli}. There were several
conference papers by Dubach et al\cite{dubach} showing results of preliminary
calculations with a full meson exchange potential; a more detailed
account of their calculations has recently become available\cite{holstein}.
Finally, for completeness we mention that a study of $\Lambda N \to NN$ based on
quark rather than meson degrees was carried out in Ref.~\cite{kisslinger}.
Their model separates the process into a long-range region, to be described
by OPE, and a short-range region, modelled by a six-quark
interaction with suitably adjusted parameters.
This idea was revived in Ref. \cite{oka}.
Results for light hypernuclei
showed that the total rates preferred a relative plus sign between
quark and OPE contributions, while the neutron to proton ratio
was better reproduced with a minus sign. In the calculation of Ref.
\cite{maltman}
both OPE and OKE (one-kaon exchange) amplitudes were considered, in
addition the the quark ones,
and significant cancellations between OPE and OKE
contributions were found.
Finally, there have been some attempts to incorporate the exchange of
the $\sigma$ and $\rho$ mesons from the point of view of a correlated
two-pion exchange\cite{itonaga,shmatikov}, where the weak vertex was
obtained through the coupling of the two pions to the $\sigma$ or
$\rho$ and intermediate N and $\Sigma$ baryon states. Both
calculations find an enhancement of the central transitions leading to
a moderate increase of the neutron to proton ratio.

The motivation for this work is two-fold.
First, in contrast to most previous investigations performed in nuclear matter
this study analyses the nonmesonic hypernuclear decay in a shell model
framework. Spectroscopic factors are employed to describe the initial
hypernuclear and final nuclear structure as well as possible. Rather than
constraining the $\Lambda$N system to only $l$=0, all possible
initial and
final relative orbital angular momenta are included. To reduce the
uncertainties regarding initial and final short-range correlations (SRC)
we use realistic $\Lambda$N and NN interactions based on the
Nijmegen
baryon-baryon potential. The nuclear structure details are thus treated with
as few approximations and ambiguities as possible. We emphasize here that our
current treatment is nonrelativistic, in contrast to our previous
works\cite{ramos91,ramos92}. As discussed in detail in
Refs. \cite{sitges,assum}, one has to pay special attention to the
effects
of short-range correlations in a relativistic treatment of
 two-body matrix elements.
Secondly, our calculations are performed in a full one-boson-exchange model
that includes not only the long-ranged pion but also contributions from the
other pseudoscalar mesons, the $\eta$ and K, as well as the vector
mesons
$\rho, \omega$ and K$^*$. Since nuclear structure uncertainties have
been
eliminated or are minimal, we can use our framework to draw conclusions
regarding the sensitivity to the
underlying weak baryon-baryon-meson couplings.
The many-body matrix elements and two-body amplitudes are evaluated in
Section II and III respectively. The form of the meson exchange
potential is presented
in Section IV for the pseudoscalar and vector mesons. Section V
discusses our calculation of the coupling constants. Our results
are given in Section VI where we discuss
the influence of each meson on the total and partial rates and the
proton asymmetry. Our conclusions are presented in Section VII.

\section{Decay rate and Asymmetry}

The nonmesonic $\Lambda N \to NN$ decay rate is given
by\cite{ramos92}
\begin{equation}
\Gamma_{nm} = \int  \frac{d^3 P}{(2\pi)^3}
 \int  \frac{d^3k}{(2\pi)^3}
\overline{\sum} (2\pi) \delta(M_H-E_R-E_1-E_2)
\mid {\cal M} \mid^2
\label{eq:rate}
\end{equation}
where ${\cal M} = \left\langle \psi_R ; {\bf P k}\ S\, M_S\ T\, M_T
\right| \hat{O}_{\Lambda N \to N N}
\left| {}_\Lambda A \right\rangle$
is the amplitude for the transition from an initial hypernuclear state to a
final state which is divided into a two-nucleon state and a residual
$(A-2)$-particle state.
The quantities
$M_H$, $E_R$, $E_1$ and $E_2$ are the mass of the hypernucleus,
the energy of the residual $(A-2)$-particle system, and the total
asymptotic energies of the
emitted nucleons, respectively.
A transformation to the center of mass
({\bf P}) and relative momentum ({\bf k}) of the two outgoing nucleons
is already implied in Eq.~(\ref{eq:rate}).
The sum $\overline{\sum}$ indicates an average
over the initial hypernucleus spin
projections, $M_J$, and sum over all quantum numbers
of the residual
($A-2$)-particle system, as well as the spin and isospin
projections of the exiting nucleons.
We follow Ref.~\cite{ramos92} by assuming a weak coupling scheme where the
isoscalar $\Lambda$ in an orbit
$\alpha_\Lambda=\{n_\Lambda,l_\Lambda,s_\Lambda,j_\Lambda,m_\Lambda\}$ couples
 to
only the ground state wave function of the nuclear $(A-1)$ core
\begin{eqnarray}
\mid{}_\Lambda A \rangle^{J_I\, M_I}_{T_I\, T_{3_I}}&=&
\mid \alpha_\Lambda \rangle \times
\mid A-1\rangle  \nonumber \\
&=& \sum_{m_\Lambda \, M_C} \langle j_\Lambda m_\Lambda
\, J_C M_C \mid J_I M_I \rangle
\mid (n_\Lambda l_\Lambda
s_\Lambda) j_\Lambda m_\Lambda \rangle
\mid J_C M_C \, T_I T_{3_I} \rangle
\ .
\label{eq:coup}
\end{eqnarray}
Employing the technique of coefficients of
fractional parentage, the core wave function is further decomposed into
a set of states where the nucleon in an orbit
$\alpha_N=\{n_N,l_N,s_N,j_N,m_N\}$ is coupled to a residual $(A-2)$-particle
state
\begin{eqnarray}
 \mid J_C M_C\, T_I T_{3_I} \rangle &=&
 \sum_{J_R T_R j_N} \langle J_C\, T_I \{ \mid J_R \, T_R, j_N \,
 t_N
  \rangle \big[ \mid J_R,\, T_R \rangle \times \mid (n_N l_N s_N)
j_N,\,  t_N \rangle \big]^{J_C M_C}_{T_I T_{3_I}} \nonumber \\
  &=&
 \sum_{J_R T_R j_N} \langle J_C\, T_I \{ \mid J_R \, T_R, j_N \,
 t_N  \rangle \nonumber \\
 &\times & \sum_{M_R\, m_N} \sum_{T_{3_R}\, t_{3_i}}
\langle J_R M_R\, j_N m_N \mid J_C M_C \rangle
\langle T_R T_{3_R}\, t_N t_{3_i} \mid T_I T_{3_I} \rangle
\nonumber \\
&\times& \mid J_R M_R \rangle \mid T_R T_{3_R} \rangle
\mid (n_N l_N s_N) j_N m_N \rangle \mid t_N t_{3_i} \rangle
\ ,
\label{eq:cfp}
\end{eqnarray}
where $t_N=1/2$. The spectroscopic factors, $S=(A-1)
\langle J_C\, T_I \{ \mid J_R \, T_R, j_N \,
t_N \rangle ^2$, appropriate for the decay of $^{12}_\Lambda$C,
are taken from Ref. \cite{cohen} and are listed in Table 4 of Ref.
\cite{ramos92}. Taking into account that the $\Lambda$ decays
from a $l_\Lambda=0$ state and working in a coupled two-body spin
and isospin basis,
the nonmesonic decay rate in
Eq.~(\ref{eq:rate}) can be written as
\begin{equation}
\Gamma_{nm}=\Gamma_{n}+\Gamma_{p} \ ,
\end{equation}
where
\begin{eqnarray}
\Gamma_{i} &=& \int \frac{d^3 P}{(2\pi)^3} \,
 \int \frac{d^3k}{(2\pi)^3}\,
(2\pi) \delta(M_H-E_R-E_1-E_2) \,
\sum_{S M_S} \sum_{J_R M_R} \sum_{T_R T_{3_R}}
\frac{1}{2J_I+1} \nonumber \\ &\times& \sum_{M_I}
\mid \langle T_R T_{3_R} \frac{1}{2} t_{3_i} \mid T_I T_{3_I}
\rangle \mid^2  \nonumber \\
&\times& \left| \,\,\sum_{T T_3}
\langle T T_3 \mid \frac{1}{2} t_1 \, \frac{1}{2} t_2 \rangle
\sum_{m_\Lambda M_C} \langle j_\Lambda m_\Lambda \, J_C M_C \mid
J_I M_I\rangle
\sum_{j_N} \langle J_C\, T_I \{ \mid J_R \, T_R, j_N \, t_N \rangle
\right. \nonumber \\
&\times&\sum_{M_R m_N} \langle J_R M_R\, j_N m_N \mid J_C M_C \rangle
 \sum_{m_{l_N} m_{s_N}} \langle j_N m_N \mid l_N m_{l_N} \frac{1}{2}
m_{s_N} \rangle
\sum_{S_0 M_{S_0}}
\langle S_0 M_{S_0} \mid \frac{1}{2} m_\Lambda \, \frac{1}{2} m_{s_N}
\rangle \nonumber \\
&\times& \left. \sum_{T_0 T_{3_0}}
\langle T_0 T_{3_0} \mid \frac{1}{2}\,\, -\frac{1}{2} \,\,\,\,
\frac{1}{2} t_{3_i} \rangle
\,\,\, t_{\Lambda N \to N N}
(S,M_S,T,M_T,S_0,M_{S_0},T_0,T_{3_0},{\bf P},{\bf k})
\,\,\right|^2  \ ,
\label{eq:rate2}
\end{eqnarray}
with $t_{3_i}=1/2$, $t_1=-1/2$, $t_2=1/2$ for the p-induced rate
($\Lambda p \to n p$) and
 $t_{3_i}=-1/2$, $t_1=-1/2$, $t_2=-1/2$ for the n-induced rate
($\Lambda n \to nn$).
Equation (\ref{eq:rate2}) is written in terms of the elementary
amplitude
$ t_{\Lambda N \rightarrow N N} $, which accounts for the transition
from an initial $\Lambda$N state with spin (isospin) $S_0$ $(T_0)$
to a final antisymmetric NN state with spin (isospin) $S$ $(T)$.
The details on how this two-body amplitude is calculated are given in the
next section.
Note that the $\Lambda$ has been assumed to act as a $\mid 1/2 \,\,
-1/2 \rangle$ isospin state which is coupled to the nucleon to total
isospin $T_0$. As explained in the next section, this is the way to
incorporate the change in isospin $\Delta I=1/2$ induced by the weak
transition operator.

At the kinematic conditions of the $(\pi^+,K^+)$ reaction carried
out at KEK, the hypernucleus is created with a substantial amount of
polarization in the ground state.
Due to the interference between the
parity conserving and parity violating amplitudes, the distribution
of the emitted protons in the weak decay displays an angular asymmetry
with respect to the polarization axis given by
\begin{equation}
\sigma(\chi)=\sigma_0 ( 1 + P_y A_p(\chi) )
\end{equation}
where $P_y$ is the hypernuclear polarization, created in the
production reaction,
such as ($\pi^+, K^+$) at KEK and BNL or ($\gamma, K^+$) at
CEBAF\cite{bennhold}, and the expression for the asymmetry
\begin{equation}
A_p(\chi)=  \frac{3}{J+1}
\frac{Tr({\cal M }
\hat{S}_y {\cal M^\dagger})}{ Tr ({\cal M M^\dagger})} =
\frac{3}{J+1}  \frac{{\displaystyle\sum_{M_i}}
\sigma (M_i)  M_i}{{\displaystyle\sum_{M_i}}  \sigma (M_i)}
\cos{\chi} = A_p \cos{\chi} \ ,
\label{eq:asympar}
\end{equation}
shown, for instance, in Ref. \cite{ramos92}, defines the asymmetry
parameter, $A_p$, characteristic of the hypernuclear weak decay
process.
The asymmetry in the distribution of protons is thus determined by the
product $P_y A_p$. In the weak coupling
scheme, simple angular momentum algebra relations
relate the hypernuclear polarization to the $\Lambda$ polarization
\begin{equation}
p_\Lambda =\left\{ \begin{array}{c c l}
-\displaystyle\frac{J}{J+1} P_y &\phantom{AAAA}& {\rm if } \ \
J=J_C-\frac{1}{2} \\
 P_y &\phantom{AAAA}& {\rm if } \ \ J=J_C+\frac{1}{2} \end{array}
\right.           \ ,
\label{eq:polar}
\end{equation}
where $J_C$ is the spin of the nuclear core.
It is convenient to introduce the intrinsic
$\Lambda$ asymmetry
parameter
\begin{equation}
a_\Lambda =\left\{ \begin{array}{c c l}
-\displaystyle\frac{J+1}{J} A_p &\phantom{AAAA}& {\rm if } \ \
J=J_C-\frac{1}{2} \\
 A_p &\phantom{AAAA}& {\rm if } \ \ J=J_C+\frac{1}{2} \end{array}
\right.           \ ,
\label{eq:asym}
\end{equation}
such that $P_y A_p=p_\Lambda
a_\Lambda$, which is then characteristic of the elementary
$\Lambda$ decay process,
$\vec{\Lambda} N \to N N$, taking place in the nuclear medium.

\section{Two-body amplitudes}

In this section we describe how we evaluate the elementary
two-body transition amplitude
$ t_{\Lambda N \rightarrow N N}$ which, as shown in
Eq.~(\ref{eq:rate2}), contains the dynamics of the weak decay
process.
In the first place, it is necessary to rewrite the product of two single
particle wave functions, $\langle {\bf r}_1 \mid \alpha_\Lambda \rangle$
and $\langle {\bf r}_2 \mid \alpha_N \rangle$, in terms of relative
and center-of-mass coordinates, ${\bf r}$ and ${\bf R}$.
In the present work, the single particle $\Lambda$ and N orbits
are taken to be solutions of harmonic oscillator mean field potentials
with
parameters $b_\Lambda=1.87$ fm and $b_N=1.64$ fm, respectively, that
have been adjusted to experimental separation energies
and the $^{12}$C charge form factor.
Assuming an average size parameter $b=(b_\Lambda + b_N)/2$,
using Moshinsky
brackets and working in the $LS$ representation,
the product of the
two harmonic oscillator single particle states,
$\Phi^{\Lambda}_{nlm}({\bf r}_1)$ and $\Phi^{N}_{n'l'm'}({\bf r}_2)$,
can be transformed to
a linear combination of products of relative and center-of-mass wave
functions. Since the $\Lambda$ is in a $l_\Lambda=0$ shell, we obtain
\begin{equation}
\Phi^{\Lambda}_{100}\left(\frac{{\bf r}_1}{b}\right)
\Phi^{N}_{100}\left(\frac{{\bf r}_2}{b}\right) =
\Phi^{\rm rel}_{100}\left(\frac{{\bf r}}{\sqrt{2}b}\right)
\Phi^{\rm CM}_{100}\left(\frac{{\bf R}}{b/\sqrt{2}}\right)  \ ,
\label{eq:sshell}
\end{equation}
when the nucleon is in the s-shell and
\begin{equation}
\Phi^{\Lambda}_{100}\left(\frac{{\bf r}_1}{b}\right)
\Phi^{N}_{11m}\left(\frac{{\bf r}_2}{b}\right) =
\frac{1}{\sqrt{2}} \left\{ \Phi^{\rm rel}_{100}\left(\frac{{\bf
r}}{\sqrt{2}b}\right)
\Phi^{\rm CM}_{11m}\left(\frac{{\bf R}}{b/\sqrt{2}}\right)  -
\Phi^{\rm rel}_{11m}\left(\frac{{\bf
r}}{\sqrt{2}b}\right)
\Phi^{\rm CM}_{100}\left(\frac{{\bf R}}{b/\sqrt{2}}\right) \right\}
\label{eq:pshell}
\end{equation}
when the nucleon is in the p-shell.

As for the final NN state,
the antisymmetric state of
two independently moving nucleons
with center-of-mass momentum ${\bf P}$ and
relative momentum ${\bf k}$ reads
\begin{equation}
\left\langle {\bf R}\, {\bf r} \right| {\bf P}\,{\bf k}\ S\, M_S\ T\, M_T
\rangle =
\frac{1}{\sqrt{2}} {\rm e}^{i{\bf P R}}\left( {\rm e}^{i{\bf k r}}
- (-1)^{S+T} {\rm e}^{-i{\bf k r}}
\right) \chi_{M_S}^S \chi_{M_T}^T  \ .
\label{eq:antwf}
\end{equation}
To incorporate the effects of the NN interaction
a substitution of the plane wave by a distorted wave
\begin{equation}
{\rm e}^{i{\bf k r}} \to \Psi_{\bf k}({\bf r}) \nonumber
\end{equation}
needs to be done.
Therefore, the amplitude
$ t_{\Lambda N \rightarrow N N}$ of Eq. (\ref{eq:rate2}) can be
decomposed in terms of amplitudes which depend on C.M. and relative
quantum numbers
\begin{eqnarray}
t_{\Lambda N \to N N} &=
& \sum_{N_r L_r N_R L_R} X(N_r L_r N_R L_R, l_\Lambda l_N)
t_{\Lambda N \to N N}^{N_r L_r\, N_R L_R} \
\end{eqnarray}
where
$X(N_r L_r N_R L_R, l_\Lambda l_N)$
are the Moshinsky brackets which for $l_\Lambda$=$l_N$=0 are just
$X(1\ 1\ 0\ 0,\ 0\ 0)=1$, and for $l_N=1$ are $X(1\ 1\ 1\ 0,\ 0\
1)=1/\sqrt{2}$ and
$X(1\ 1\ 0\ 1,\ 0\ 1)=-1/\sqrt{2}$, as can be seen from the
decomposition of the wave function above. The matrix elements
$t_{\Lambda N \to N N}^{N_r L_r\, N_R L_R}$
are given by
\begin{eqnarray}
t_{\Lambda N \to N N}^{N_r L_r\, N_R L_R} &=
& \frac{1}{\sqrt{2}}\int d^3R \int d^3r \, {\rm e}^{-i{\bf PR}}
\Psi^*_{\bf k} ({\bf r}) \chi^{\dagger\, S}_{M_S}
\chi^{\dagger\, T}_{T_3} \, V({\bf r}) \,
\Phi^{\rm CM}_{N_R L_R}\left(\frac{{\bf R}}{b/\sqrt{2}}\right)
\Phi^{\rm rel}_{N_r L_r}\left(\frac{{\bf r}}{\sqrt{2}b}\right)
\chi^{S_0}_{M_{S_0}} \chi^{T_0}_{T_{3_0}}\nonumber \\
&=& \Phi^{CM}_{N_R L_R} \left({\bf K}\frac{b}{\sqrt{2}}\right)
\,\, t_{\rm rel} \ ,
\label{eq:two}
\end{eqnarray}
with
\begin{equation}
t_{\rm rel}= \frac{1}{\sqrt{2}}
 \int d^3r \,
\Psi^*_{\bf k} ({\bf r}) \chi^{\dagger\, S}_{M_S}
\chi^{\dagger\, T}_{T_3} \, V({\bf r}) \,
\Phi^{\rm rel}_{N_r L_r} \left(\frac{{\bf r}}{\sqrt{2}b}\right)
\chi^{S_0}_{M_{S_0}} \chi^{T_0}_{T_{3_0}} \ ,
\label{eq:trel}
\end{equation}
where, for simplicity, we have only shown the direct amplitude
corresponding to the first term of Eq. (\ref{eq:antwf}).
The function $\Phi^{CM}_{N_R L_R}\left({\bf K}
\displaystyle\frac{b}{\sqrt{2}}\right)$ is the Fourier transform of
the $\Lambda$N center-of-mass wave function and
$t_{\rm rel}$ is the expectation value of the transition
potential $V({\bf r})$ between $\Lambda$N and NN
relative wave functions.
In the next section it is shown how the potential $V({\bf r})$
can be decomposed as
\begin{equation}
V({\bf r})=\sum_i \sum_\alpha V_\alpha^{(i)}(r)
\hat{O}_\alpha \hat{I}_\alpha^{(i)}
\ ,
\end{equation}
where the index $i$ runs over the different mesons exchanged and
$\alpha$ over the different spin operators, $\hat{O}_\alpha \in
(\hat{1},
\ \mbox{\boldmath $\sigma$}_1
\mbox{\boldmath $\sigma$}_2,
\ S_{12}(\hat{\bf r})=
3\mbox{\boldmath $\sigma$}_1 \hat{\bf r}
\mbox{\boldmath $\sigma$}_2 \hat{\bf r} -
\mbox{\boldmath $\sigma$}_1
\mbox{\boldmath $\sigma$}_2,
\ \mbox{\boldmath $\sigma$}_2 \hat{\bf r},
\ [\mbox{\boldmath $\sigma$}_1 \times
\mbox{\boldmath $\sigma$}_2] \hat{\bf r})$,
which occur in
the potential $V(\bf r)$. The isospin operator,
$\hat{I}_\alpha^{(i)}$, depends on
the meson and can be either $\hat{1}$ for isoscalar mesons ($\eta$,
$\omega$),
$\mbox{\boldmath $\tau$}_1
\mbox{\boldmath $\tau$}_2$ for isovector mesons ($\pi$, $\rho$) or a
linear combination of
$\hat{1}$ and $\mbox{\boldmath $\tau$}_1
\mbox{\boldmath $\tau$}_2$, with the coefficients depending on the
particular spin structure piece of the potential, for isodoublet
mesons (K, K$^*$).
The radial parts
$V^{(i)}_\alpha(r)$ are discussed in the next section.

By performing a partial wave expansion of
the final two-nucleon wave function and working in the
$(LS)J$-coupling scheme,
the relative $\Lambda N\to NN$
amplitude, $t_{\rm rel}$, can be further decomposed

\begin{eqnarray}
t_{\rm rel} &=&  \frac{1}{\sqrt{2}} \sum_i\sum_\alpha \sum_{L L' J}
4\pi i^{-L'}
\langle L M_{L} S M_{S} | J M_J \rangle Y_{L M_L} (\hat{\bf k}_r)
\langle L_{r} M_{L_{r}} S_{0} M_{S_{0}} | J M_J \rangle
\nonumber \\
&\times &
\langle (L' S) J M_J\mid \hat{O}_\alpha \mid (L_r S_0)  J M_J\rangle
\langle T T_3 \mid \hat{I}_\alpha^{(i)} \mid T_0 T_{3_0}\rangle
\int r^2 dr \, \Psi^{*\,J}_{L L'} (k_r,r) V^{(i)}_\alpha(r)
\Phi^{\rm rel}_{N_r L_r} (r) \ ,
\label{eq:trel2}
\end{eqnarray}
where $\Phi^{\rm rel}_{N_r L_r}(r)$ stands for the radial piece of
the H.O. wave function. The explicit expressions for the expectation
value of the spin-space piece,
$\langle (L' S) J M_J\mid \hat{O}_\alpha \mid (L_r S_0)  J
M_J\rangle$,
can be found in the Appendix.

The function $\Psi^{\ J}_{L L'} (k_r,r)$ is the scattering solution of
two nucleons moving under
the influence of the strong interaction, for which we consider
the updated version of the Reid-soft core potential\cite{reid}, given in
Ref. \cite{stoks}, and the Nijmegen\cite{stoks} NN
potential. Such a wave function is obtained by solving
a $T$-matrix equation in momentum space and in partial wave
decomposition following the method described in Ref. \cite{haftel}.
The tensor component of the NN interaction couples relative orbital
states ($L$ and $L'$) having the same parity and total angular
momentum as, for instance,
the $^3S_1$ and $^3D_1$ channels. Therefore, starting from an initial
$L_r$ orbital momentum, the weak transition potential produces a
transition to a $L'$ value, which mixes, through the subsequent action
of the strong interaction, with another value of orbital angular momentum,
$L$. In Table \ref{tab:chan} we present all the possible final
states
starting from initial $\Lambda$N states having either $L_r=0$ or
1 and for the central ($\Delta {\bf S}=0$, $\Delta {\bf L}=0$),
tensor ($\Delta {\bf S}=2$, $\Delta {\bf L}=2$)
and parity violating ($\Delta {\bf S}=1$, $\Delta {\bf L}=1$)
pieces in which the transition potential can be decomposed.

In the absence of final state interactions (FSI), the NN wave
function
in Eq. (\ref{eq:trel2}) would reduce to a spherical Bessel function
\begin{equation}
\Psi^{\ J}_{ L L'} (k_r,r) = \delta_{L,L'}
j_{L} (k_r r) \ .
\end{equation}
We note that the procedure followed here to include
FSI between the two emitted nucleons differs from our previous
works\cite{rho,praga}, where the non-interacting NN pair,
represented by a Bessel function
in the final state, was multiplied by
an average NN correlation function
\begin{equation}
f_{\rm FSI}(r)=1-j_0(q_c r),
\end{equation}
with $q_c=3.93$ fm$^{-1}$,
which provides a good description of nucleon pairs in $^4$He
\cite{weise} as calculated with the Reid-soft core interaction
\cite{reid}.

To account for the $\Lambda$N correlations, which are absent in the
independent particle model, one should replace the harmonic oscillator
$\Lambda$N wave function, $\Phi^{\rm rel}_{N_r L_r}(r)$, by a
correlated
$\Lambda$N wave function that contains the effect of the
strong $\Lambda$N interaction. Such wave functions were obtained
from a microscopic finite-nucleus $G$-matrix
calculation\cite{halder} using the soft-core and hard-core
Nijmegen models\cite{nijme}. In Ref. \cite{sitges} we showed that
multiplying the uncorrelated harmonic oscillator
$\Lambda$N wave function with the spin-independent
correlation function
\begin{equation}
f(r)=\left( 1 - {\rm e}^{-r^2 / a^2} \right)^n + b r^2 {\rm
e}^{-r^2 / c^2} \ ,
\label{eq:cor}
\end{equation}
with $a=0.5 $, $b=0.25 $, $c= 1.28$, $n= 2$,
yielded decay rates slightly larger than those obtained with the
numerical Nijmegen soft-core correlations but slightly smaller than
those computed with the Nijmegen hard-core potential. Since the
deviations were at most 10\% the above parametrization can be used as a
good approximation to the full correlation function.

\section{The Meson Exchange Potential}
\begin{figure}
       \setlength{\unitlength}{1mm}
       \begin{picture}(100,180)
      \put(25,0){\epsfxsize=12cm \epsfbox{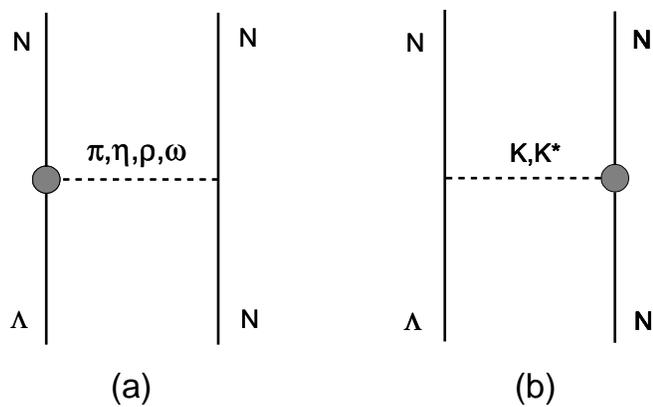}}
       \end{picture}
\caption{
Non-strange (a) and strange (b) meson exchange contribution to the
$\Lambda N \to N N$ weak transition potential.
The weak vertex is indicated by the circle.
}
\label{fig:amp}
\end{figure}

The transition $\Lambda N \to N N$ is assumed to proceed via
the exchange of virtual mesons belonging to the ground-state
pseudoscalar and vector meson octets. As displayed in Figs.
\ref{fig:amp}(a) and
(b), the transition amplitude
involves a strong and a weak vertex, the later being denoted by a
hatched circle.

\subsection{Pseudoscalar mesons}

While there exist
several strong meson-exchange potentials which, through fits
to NN scattering data, provide information
on the different strong NN-meson vertices, only the pion vertex is
known experimentally in the weak sector. The weak hamiltonian is
parametrized in the form
\begin{equation}
{\cal H}^{\rm W }_{\rm {\scriptscriptstyle \Lambda N} \pi}= i G_F m_\pi^2
\overline{\psi}_{\rm N}
(A_\pi+B_\pi \gamma_5)
\mbox{\boldmath $\tau \phi$}^\pi
\psi_\Lambda \, \left( ^0_1 \right) \ ,
\label{eq:weak}
\end{equation}
where $G_F m_\pi^2= 2.21\times 10^{-7}$ is the weak coupling constant.
The empirical constants
$A_\pi=1.05$ and $B_\pi=-7.15$, adjusted to the observables of the
free
$\Lambda$ decay, determine
the strength of the parity violating and parity conserving
amplitudes, respectively.
The nucleon, lambda and pion fields are given by
$\psi_{\rm N}$, $\psi_\Lambda$ and $\mbox{\boldmath $\phi$}^\pi$,
respectively, while
the isospin spurion $\left( ^0_1 \right)$ is included
to enforce the empirical
$\Delta I=1/2$ rule observed in the decay of a free
$\Lambda$.

For the strong vertex, we take the usual pseudoscalar coupling
\begin{equation}
{\cal H}^{\rm S }_{\rm {\scriptscriptstyle NN} \pi}= i g_{\rm
 {\scriptscriptstyle NN} \pi}
\overline{\psi}_{\rm N}
\gamma_5
\mbox{\boldmath{$\tau$}{$\phi^\pi$}} \psi_{\rm N} \ ,
\label{eq:strongpi}
\end{equation}
which
is equivalent to the pseudovector coupling when free spinors
are used in the evaluation of the transition amplitude.
The nonrelativistic reduction
of the free space Feynman amplitude is then associated with
the transition potential. In momentum space, one obtains
\begin{equation}
V_{\pi}({\bf q}) = - G_F m_\pi^2
\frac{g}{2M} \left(
{\hat A} + \frac{{\hat B}}{2\overline{M}}
\mbox{\boldmath $\sigma$}_1 {\bf q} \right)
\frac{\mbox{\boldmath $\sigma$}_2 {\bf q} }{{\bf q}^2+\mu^2}
\label{eq:pion}
\end{equation}
where ${\bf q}$ is the momentum carried by the pion directed towards the
strong vertex, $g = g_{\rm {\scriptscriptstyle NN}
\pi}$ the strong coupling constant for the NN$\pi$ vertex, $\mu$
the pion mass,
$M$ the nucleon mass and $\overline M$
the average between the nucleon and $\Lambda$ masses.
The operators
${\hat A}$ and ${\hat B}$, which contain the isospin
dependence of the potential, read
\begin{eqnarray}
{\hat A} &=& A_\pi \,\, \mbox{\boldmath $\tau$}_1
\mbox{\boldmath $\tau$}_2 \nonumber \\
{\hat B} &=& B_\pi \,\, \mbox{\boldmath $\tau$}_1
\mbox{\boldmath $\tau$}_2
\end{eqnarray}

We note that the nonrelativistic approach of the present work
differs from our previous works \cite{ramos91,ramos92}, which were
based on a relativistic formalism. It was found that the
supression of the matrix elements due to short range correlations
was larger by about a factor of two to what was obtained in
standard nonrelativistic calculations
\cite{salcedo,mckellar,dubach,holstein}. In Ref. \cite{sitges}
it was shown that, if one uses the same nonrelativistic correlation function,
the relativistic and nonrelativistic schemes were not giving
the same correlated potential obtained through the standard
nonrelativistic reduction.
In the relativistic approach,
the correlation function was applied to the
Feynman amplitude $\rm before$ the nonrelativistic reduction was carried
out, whereas in the
nonrelativistic procedure the correlation function was applied
$\rm after$ the reduction of the free
Feynman amplitude was obtained. The difference between the two
methods was studied in Ref. \cite{assum}, where it was
shown explicitly that the relativistic framework together with a
standard nonrelativistic correlation function lead to additional
contributions in the correlated transition potential which produced
the larger supression of the decay rates reported in Refs.
\cite{ramos91,ramos92}.
For lack of relativistic
correlation functions we adopt a nonrelativistic formalism in this
paper.

The other mesons of the pseudoscalar octet are the
isosinglet eta ($\eta$) and the
isodoublet kaon
(K). The strong and weak vertices for these mesons are
\begin{eqnarray}
{\cal H}^{\rm S }_{\rm {\scriptscriptstyle NN} \eta}&=&
 i g_{\rm {\scriptscriptstyle NN} \eta} \overline {\psi}_{\rm N} \gamma_5
\phi^\eta \psi_{\rm N}  \\
{\cal H}^{\rm W }_{\rm {\scriptscriptstyle \Lambda N} \eta}&=&
i G_F m_\pi^2 \overline{\psi}_{\rm N} \,\,
(A_\eta+B_\eta \gamma_5)
\phi^\eta \psi_\Lambda \, \left( ^0_1 \right)  \\
{\cal H}^{\rm S }_{\rm \scriptscriptstyle{\Lambda N K}}&=&
i g_{\rm {\scriptscriptstyle \Lambda N K}} \overline{\psi}_{\rm N} \gamma_5 \,\,
\phi^K \psi_\Lambda
\label{eq:kastrong} \\
{\cal H}^{\rm W }_{\rm \scriptscriptstyle{NNK}}&=&
i G_F m_\pi^2 \left[ \overline{\psi}_{\rm N} \left( ^0_1 \right)
\,\,( C_{\rm \scriptscriptstyle{K}}^{\rm \scriptscriptstyle{PV}} + C_{\rm
 \scriptscriptstyle{K}}^{\rm \scriptscriptstyle{PC}}
\gamma_5)
\,\,(\phi^K)^\dagger
\psi_{\rm N}
+ \overline{\psi}_{\rm N} \psi_{\rm N}
\,\,( D_{\rm \scriptscriptstyle{K}}^{\rm\scriptscriptstyle{PV}} +
 D_{\rm\scriptscriptstyle{K}}^{\rm\scriptscriptstyle{PC}}
\gamma_5 ) \,\,(\phi^K)^\dagger \,\,
\left( ^0_1 \right) \right]
\label{eq:kaweak}
\end{eqnarray}
where the weak coupling constants cannot be derived from experiment.
In the present work we adopt the approach of
Refs.\cite{holstein,delatorre}, presented in the next section.

\begin{figure}
       \setlength{\unitlength}{1mm}
       \begin{picture}(100,180)
      \put(25,0){\epsfxsize=12cm \epsfbox{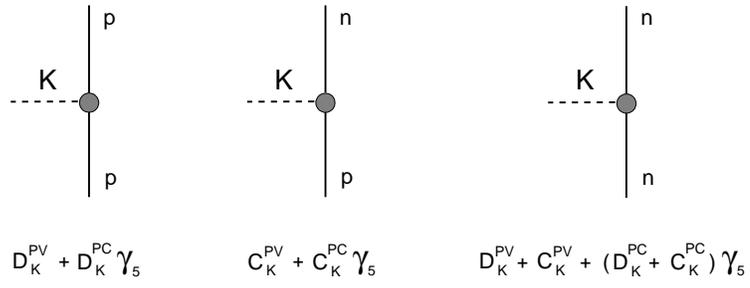}}
       \end{picture}
\caption{
K-meson weak vertices for $p\overline{p}K^0$ (a),
$p\overline{n}K^+$ (b), and
$n\overline{n}K^0$ (c).
}
\label{fig:kaon}
\end{figure}

We note that the isospurion $\displaystyle{\left(^0_1\right)}$
appearing in the former equations is used to enforce the empirical
$\Delta I=1/2$ rule.
The particular structure of the K weak couplings reproduce the
vertices shown in Fig. \ref{fig:kaon}.

The corresponding nonrelativistic potentials for the transition
$\Lambda N \to NN$ are analogous to Eq. (\ref{eq:pion})
but making the following replacements
\begin{eqnarray}
g &\to&  g_{\rm {\scriptscriptstyle NN} \eta} \nonumber \\
\mu &\to& m_\eta \nonumber \\
{\hat A} &\to& A_\eta \nonumber \\
{\hat B} &\to& B_\eta \ ,
\end{eqnarray}
in the case of $\eta$-exchange, and
\begin{eqnarray}
g &\to& g_{\rm{\scriptscriptstyle \Lambda N K}} \nonumber \\
\mu &\to& m_{\rm {\scriptscriptstyle K}} \nonumber \\
{\hat A} &\to& \left( \frac{
C^{\rm\scriptscriptstyle{PV}}_{\rm\scriptscriptstyle{K}}}{2} +
D^{\rm\scriptscriptstyle{PV}}_{\rm\scriptscriptstyle{K}} + \frac{
C^{\rm\scriptscriptstyle{PV}}_{\rm\scriptscriptstyle{K}}}{2}
\mbox{\boldmath $\tau$}_1
\mbox{\boldmath $\tau$}_2 \right) \frac{M}{\overline M}
\nonumber \\
{\hat B} &\to& \left( \frac{
C^{\rm\scriptscriptstyle{PC}}_{\rm\scriptscriptstyle{K}}}{2} +
D^{\rm\scriptscriptstyle{PC}}_{\rm\scriptscriptstyle{K}} + \frac{
C^{\rm\scriptscriptstyle{PC}}_{\rm\scriptscriptstyle{K}}}{2}
\mbox{\boldmath $\tau$}_1
\mbox{\boldmath $\tau$}_2 \right) \ ,
\end{eqnarray}
in the case of K-exchange,
where the factor $\frac{M}{\overline M}$
corrects for the fact that the nonrelativistic reduction of the
strong $\Lambda {\rm N K}$ vertex gives a factor
$1/\overline{M}$ instead of $1/M$.
Performing a Fourier transform of the general expression given
in Eq. (\ref{eq:pion}) and introducing the tensor operator
$S_{12}(\hat{\bf r}) = 3
\mbox{\boldmath $\sigma$}_1 {\hat{\bf r}}
\mbox{\boldmath $\sigma$}_2 {\hat{\bf r}} -
\mbox{\boldmath $\sigma$}_1
\mbox{\boldmath $\sigma$}_2$,
it is easy to obtain the corresponding transition potential in
coordinate
space, which can be divided into  central, tensor and parity violating
pieces. The explicit expressions are given at the end of this section.

\subsection{Vector mesons}

A number of theoretical studies in recent years have investigated the
contribution of the $\rho$-meson to the $\Lambda N \to NN$
process\cite{mckellar,nardulli,takeuchi}.
The weak $\Lambda {\rm N} \rho$ and strong NN$\rho$ vertices are given
by\cite{mckellar}
\begin{eqnarray}
{\cal H}^{\rm W}_{\rm {\scriptscriptstyle \Lambda N} \rho} &=& G_F m_\pi^2 \:
{\overline \psi}_{\rm N} \: \left( \alpha_\rho \gamma^\mu
- \beta_\rho i \frac{\sigma^{\mu \nu} q_\nu} {2 \overline{M}} +
\varepsilon_\rho
\gamma^\mu \gamma_5 \right)
\mbox{\boldmath $\tau \rho$}_\mu \psi_\Lambda \,\, \left( ^0_1
\right) \\
{\cal H}^{\rm S}_{\rm {\scriptscriptstyle NN} \rho} &=& \overline{\psi}_{\rm N}
\left( g^{\rm {\scriptscriptstyle V}}_{\rm {\scriptscriptstyle NN} \rho}
 \gamma^\mu + i
\frac{ g^{\rm {\scriptscriptstyle T}}_{\rm {\scriptscriptstyle NN} \rho}}{2M}
\sigma^{\mu \nu} q_\nu \right)
\mbox{\boldmath $\tau \rho$}_\mu \psi_{\rm N}   \ ,
\label{eq:rhohamil}
\end{eqnarray}
respectively, where the four momentum transfer $q$
is directed towards the strong vertex.
The values of the strong and weak
coupling constants are given in the next section.

The nonrelativistic reduction of the Feynman amplitude gives
the following $\rho$-meson transition potential
\begin{eqnarray}
{V_{\rho}}({\bf q})  &=&
G_F m_\pi^2
 \left( F_1 {\hat \alpha} - \frac{({\hat \alpha} + {\hat \beta} )
 ( F_1 + F_2 )} {4M \overline{M}}
(\mbox{\boldmath $\sigma$}_1 \times {\bf q})
(\mbox{\boldmath $\sigma$}_2 \times {\bf q}) \right. \nonumber \\
& & \phantom { G_F m_\pi^2 A }
\left. +i \frac{{\hat \varepsilon} ( F_1 + F_2 )} {2M}
(\mbox{\boldmath $\sigma$}_1 \times
\mbox{\boldmath $\sigma$}_2 ) {\bf q}\right)
\frac{1}{{\bf q}^2 + \mu^2} \
\label{eq:rhopot}
\end{eqnarray}
with $\mu = m_\rho$, $F_1 = g^{\rm {\scriptscriptstyle V}}_{\rm
 {\scriptscriptstyle NN} \rho}$,
$F_2 = g^{\rm {\scriptscriptstyle T}}_{\rm {\scriptscriptstyle  NN} \rho}$
and the operators ${\hat \alpha}$, ${\hat
\beta}$ and ${\hat \varepsilon}$
\begin{eqnarray}
{\hat \alpha} &=& \alpha_\rho \,\, \mbox{\boldmath $\tau$}_1
\mbox{\boldmath $\tau$}_2  \nonumber \\
{\hat \beta} &=& \beta_\rho \,\, \mbox{\boldmath $\tau$}_1
\mbox{\boldmath $\tau$}_2  \nonumber \\
{\hat \varepsilon} &=& \varepsilon_\rho \,\, \mbox{\boldmath
$\tau$}_1
\mbox{\boldmath $\tau$}_2
\end{eqnarray}
contain the isospin structure.
Using the relation
$({\mbox{\boldmath $\sigma$}}_1 \times {\bf q}) ({\mbox{\boldmath $\sigma$}}_2
\times {\bf q}) =
({\mbox{\boldmath $\sigma$}}_1 {\mbox{\boldmath $\sigma$}}_2)
\: {\bf q}^2 - ({\mbox{\boldmath $\sigma$}}_1 {\bf q})
 ({\mbox{\boldmath $\sigma$}}_2 {\bf q})$
and performing a Fourier transform of $V_{\rho} ({\bf q})$,
one obtains the corresponding transition potential in coordinate
space, which, as in $\pi$ exchange, can be divided into
central, tensor and parity-violating pieces.
Furthermore, the $\rho$-meson central potential can be further
decomposed into a spin independent and a spin dependent
part\cite{rho}.
Due to the different models employed for the weak
$\Lambda {\rm N} \rho$-vertex\cite{mckellar,nardulli,takeuchi},
different calculations have yielded widely varying results.  However,
all studies until now have only
included the tensor piece of the parity-conserving
$\rho$-exchange term motivated, in part, by the observation
that this is the most important contribution to the
$\pi$-exchange potential.
We recently demonstrated\cite{rho} that
the central piece of the
$\rho$-exchange is in fact larger than
its tensor interaction, an observation that can be traced
to the fact that the
$\rho$-exchange diagram has a much shorter range than the $\pi$-exchange
potential. It is therefore important to explicitly keep all pieces of
the potential for the vector mesons.

The other vector mesons considered in this work are
the isoscalar $\omega$ and the isodoublet K$^*$, for which the
weak and strong vertices can be written as
\begin{eqnarray}
{\cal H}^{\rm S }_{\rm {\scriptscriptstyle NN} \omega}&=&
\overline{\psi}_{\rm N} \left( g^{\rm {\scriptscriptstyle V}}_{\rm
 {\scriptscriptstyle NN} \omega}
\gamma^\mu + i \frac{g^{\rm {\scriptscriptstyle T}}_{\rm {\scriptscriptstyle NN}
 \omega}}{2M}
\sigma^{\mu \nu} q_\nu \right)
\phi^\omega_\mu \psi_{\rm N} \\
{\cal H}^{\rm W}_{\rm {\scriptscriptstyle \Lambda N} \omega} &=& G_F m_\pi^2 \:
{\overline \psi}_{\rm N}
\:
\left( \alpha_\omega \gamma^\mu
- \beta_\omega i \frac{\sigma^{\mu \nu} q_\nu} {2 \overline M} +
\varepsilon_\omega
\gamma^\mu \gamma_5 \right)
\phi^\omega_\mu \psi_\Lambda \left( ^0_1 \right)
 \\
{\cal H}^{\rm S }_{\rm {\scriptscriptstyle \Lambda N K^*}}&=&
\overline{\psi}_{\rm N} \left(g^{\rm \scriptscriptstyle{V}}_{\rm
 \scriptscriptstyle{\Lambda N
K^*}}
\gamma^\mu + i \frac{g^{\rm \scriptscriptstyle{V}}_{\rm
 \scriptscriptstyle{\Lambda N K^*}}}{2
\overline M}
\sigma^{\mu \nu} q_\nu \right)
\phi^{\rm\scriptscriptstyle{K^*}}_\mu \psi_\Lambda \\
{\cal H}^{\rm W}_{\rm\scriptscriptstyle{NN K^*}} &=& G_F m_\pi^2 \: \left( \,\,
\left[C_{\rm\scriptscriptstyle{K^*}}^{\rm \scriptscriptstyle{PC,V}}
 \overline{\psi}_{\rm N} \left(
^0_1 \right)
(\phi^{\rm\scriptscriptstyle{K^*}}_\mu)^\dagger\,\,
\gamma^\mu \psi_{\rm N} +
D_{\rm\scriptscriptstyle{K^*}}^{\rm\scriptscriptstyle{PC,V}}
 \overline{\psi}_{\rm N} \gamma^\mu
\psi_{\rm N}
(\phi^{\rm\scriptscriptstyle{K^*}}_\mu)^\dagger \,\,
\left( ^0_1 \right) \right] + \right. \nonumber \\
& & \left. \left[ C_{\rm\scriptscriptstyle{K^*}}^{\rm\scriptscriptstyle{PC,T}}
\overline{\psi}_{\rm N}
\left( ^0_1 \right) (\phi^{\rm\scriptscriptstyle{K^*}}_\mu)^\dagger\,\,
(-i) \frac{\sigma^{\mu \nu} q_\nu} {2 M}
\psi_{\rm N} +
D_{\rm\scriptscriptstyle{K^*}}^{\rm\scriptscriptstyle{PC,T}}
 \overline{\psi}_{\rm N}
(-i) \frac{\sigma^{\mu \nu} q_\nu} {2 M}
\psi_{\rm N}
(\phi^{\rm\scriptscriptstyle{K^*}}_\mu)^\dagger \,\,
\left( ^0_1 \right) \right] +  \right. \nonumber \\
& & \left.
\left[ C_{\rm\scriptscriptstyle{K^*}}^{\rm\scriptscriptstyle{PV}}
 \overline{\psi}_{\rm N} \left(
^0_1 \right)
(\phi^{\rm\scriptscriptstyle{K^*}}_\mu)^\dagger\,\,
\gamma^\mu \gamma_5 \psi_{\rm N} +
D_{\rm\scriptscriptstyle{K^*}}^{\rm\scriptscriptstyle{PV}} \overline{\psi}_{\rm
 N} \gamma^\mu
\gamma_5 \psi_{\rm N}
(\phi^{\rm\scriptscriptstyle{K^*}}_\mu)^\dagger \,\,
\left( ^0_1 \right) \right] \,\, \right)
\end{eqnarray}
Note that the K$^*$ weak vertex has the same structure as the K
vertex, the only difference being the parity conserving
contribution which has two terms, related to the vector
and tensor couplings. The nonrelativistic potential
can be obtained from the general expression given in
Eq. (\ref{eq:rhopot})
making the following replacements
\begin{eqnarray}
\mu &\to& m_\omega \nonumber \\
F_1 &\to& g^{\rm {\scriptscriptstyle V}}_{\rm {\scriptscriptstyle NN} \omega}
 \nonumber \\
F_2 &\to& g^{\rm {\scriptscriptstyle T}}_{\rm {\scriptscriptstyle NN} \omega}
 \nonumber \\
{\hat \alpha} &\to& \alpha_\omega \nonumber \\
{\hat \beta} &\to& \beta_\omega \nonumber \\
{\hat \varepsilon} &\to& \varepsilon_\omega \ ,
\end{eqnarray}
in the case of $\omega$ exchange, and
\begin{eqnarray}
\mu &\to& m_{\rm {\scriptscriptstyle K^*}} \nonumber \\
F_1 &\to& g^{\rm \scriptscriptstyle{V}}_{\rm {\scriptscriptstyle \Lambda N K^*}}
 \nonumber \\
F_2 &\to& g^{\rm \scriptscriptstyle{T}}_{\rm {\scriptscriptstyle \Lambda N K^*}}
 \nonumber \\
{\hat \alpha} &\to& \frac{
C^{\rm \scriptscriptstyle{PC, V}}_{\rm \scriptscriptstyle{K^*}}} {2} +
D^{\rm \scriptscriptstyle{PC, V}}_{\rm \scriptscriptstyle{K^*}} + \frac{
C^{\rm \scriptscriptstyle{PC, V}}_{\rm \scriptscriptstyle{K^*}}} {2}
\mbox{\boldmath $\tau$}_1
\mbox{\boldmath $\tau$}_2
\nonumber \\
{\hat \beta} &\to& \frac{
C^{\rm \scriptscriptstyle{PC, T}}_{\rm \scriptscriptstyle{K^*}}} {2} +
D^{\rm \scriptscriptstyle{PC, T}}_{\rm \scriptscriptstyle{K^*}} + \frac{
C^{\rm \scriptscriptstyle{PC, T}}_{\rm \scriptscriptstyle{K^*}}} {2}
\mbox{\boldmath $\tau$}_1
\mbox{\boldmath $\tau$}_2
\nonumber \\
{\hat \varepsilon} &\to& \left( \frac {
C^{\rm \scriptscriptstyle{PV}}_{\rm \scriptscriptstyle{K^*}}} {2} +
D^{\rm \scriptscriptstyle{PV}}_{\rm \scriptscriptstyle{K^*}} + \frac{
C^{\rm \scriptscriptstyle{PV}}_{\rm \scriptscriptstyle{K^*}}} {2}
\mbox{\boldmath $\tau$}_1
\mbox{\boldmath $\tau$}_2 \right) \frac{M}{\overline M} \ ,
\label{kstar}
\end{eqnarray}
for the K$^*$ meson exchange.

\subsection{General Form of the potential}

The Fourier transform of the general Eqs. (\ref{eq:pion}) and
(\ref{eq:rhopot}) leads to a potential in configuration space which
can be cast into the form
\begin{eqnarray}
V({\bf r}) &=& \sum_{i} \sum_\alpha V_\alpha^{(i)}
({\bf r}) = \sum_i \sum_{\alpha}
V_\alpha^{(i)} (r) \hat{O}_\alpha \hat{I}_\alpha^{(i)} \nonumber
\\
&=& \sum_{i} \left[ V_C^{(i)}(r) \hat{I}^{(i)}_C + V_{SS}^{(i)}(r)
\mbox{\boldmath
$\sigma$}_1
\mbox{\boldmath $\sigma$}_2 \hat{I}^{(i)}_{SS}
+ V_T^{(i)}(r)
S_{12}(\hat{\bf r}) \hat{I}^{(i)}_T + \right. \nonumber \\
& & \left. + \left( n^i \mbox{\boldmath $\sigma$}_2
\cdot \hat{\bf r}
+ (1-n^i) \left[\mbox{\boldmath $\sigma$}_1 \times
\mbox{\boldmath
$\sigma$}_2 \right] \cdot \hat{\bf r} \right)
V_{PV}^{(i)}(r) \hat{I}^{(i)}_{PV} \right] \ ,
\label{eq:genpot}
\end{eqnarray}
where the index $i$ runs over the different mesons exchanged ($i=1,\dots,
6$ represents $\pi,\rho$,K,K$^*$,$\eta,\omega$)
and
$\alpha$ over the different
spin operators denoted by $C$
(central spin independent), $SS$ (central spin dependent), $T$
(tensor) and $PV$ (parity violating). In the above expression,
particle 1 refers to the
$\Lambda$ and
$n^i = 1 (0)$ for pseudoscalar (vector) mesons.
In the case of isovector mesons ($\pi$, $\rho$) the
isospin factor is
$\mbox{\boldmath $\tau$}_1
\mbox{\boldmath $\tau$}_2$ and for
isoscalar mesons
($\eta$,$\omega$)
this factor is just $\hat{1}$ for all spin structure pieces of the potential.
In the case of isodoublet mesons (K,
K$^*$) there
are contributions proportional to $\hat{1}$ and to $\mbox{\boldmath
$\tau$}_1 \mbox{\boldmath $\tau$}_2$ that depend on the
coupling constants and, therefore, on the
spin structure piece of the potential denoted by $\alpha$.
For K-exchange we have
\begin{eqnarray}
{\hat I}_{C}^{(3)}&=& 0 \nonumber \\
{\hat I}_{SS}^{(3)}&=& {\hat I}_{T}^{(3)} =
\frac{
C^{\rm \scriptscriptstyle{P C}}_{\rm\scriptscriptstyle{K}}} {2} +
D^{\rm \scriptscriptstyle{P C}}_{\rm\scriptscriptstyle{K}} + \frac{
C^{\rm \scriptscriptstyle{P C}}_{\rm\scriptscriptstyle{K}}} {2}
\mbox{\boldmath $\tau$}_1
\mbox{\boldmath $\tau$}_2 \nonumber \\
{\hat I}_{PV}^{(3)} &=&
\frac{
C^{\rm \scriptscriptstyle{P V}}_{\rm\scriptscriptstyle{K}}} {2} +
D^{\rm \scriptscriptstyle{P V}}_{\rm\scriptscriptstyle{K}} + \frac{
C^{\rm \scriptscriptstyle{P V}}_{\rm\scriptscriptstyle{K}}} {2}
\mbox{\boldmath $\tau$}_1
\mbox{\boldmath $\tau$}_2
\end{eqnarray}
and for K$^*$ exchange
\begin{eqnarray}
{\hat I}_{C}^{(6)} &=&  \frac{
C^{\rm \scriptscriptstyle{PC,V}}_{\rm\scriptscriptstyle{K^*}}} {2} +
D^{\rm \scriptscriptstyle{PC,V}}_{\rm\scriptscriptstyle{K^*}} + \frac{
C^{\rm \scriptscriptstyle{PC,V}}_{\rm\scriptscriptstyle{K^*}}} {2}
\mbox{\boldmath $\tau$}_1
\mbox{\boldmath $\tau$}_2 \nonumber \\
{\hat I}_{SS}^{(6)} &=& {\hat I}_{T}^{(6)} =
\frac { \left(
C^{\rm \scriptscriptstyle{PC,V}}_{\rm\scriptscriptstyle{K^*}} +
C^{\rm \scriptscriptstyle{PC,T}}_{\rm\scriptscriptstyle{K^*}} \right)} {2}
+ \left( D^{\rm \scriptscriptstyle{PC,V}}_{\rm\scriptscriptstyle{K^*}} +
D^{\rm \scriptscriptstyle{PC,T}}_{\rm\scriptscriptstyle{K^*}} \right) +
\frac{ \left( C^{\rm \scriptscriptstyle{PC,V}}_{\rm\scriptscriptstyle{K^*}} +
C^{\rm \scriptscriptstyle{PC,T}}_{\rm\scriptscriptstyle{K^*}} \right) }{2}
\mbox{\boldmath $\tau$}_1
\mbox{\boldmath $\tau$}_2 \nonumber \\
{\hat I}_{PV}^{(6)} &=&
\frac{
C^{\rm \scriptscriptstyle{PV}}_{\rm\scriptscriptstyle{K^*}}} {2} +
D^{\rm \scriptscriptstyle{PV}}_{\rm\scriptscriptstyle{K^*}} + \frac {
C^{\rm \scriptscriptstyle{PV}}_{\rm\scriptscriptstyle{K^*}}} {2}
\mbox{\boldmath $\tau$}_1
\mbox{\boldmath $\tau$}_2  \ .
\end{eqnarray}

The different pieces $V_{\alpha}^{(i)}$, with $\alpha=C,SS,T,PV$,
are given by
\begin{eqnarray}
V_{C}^{(i)} (r) &=&  K^{(i)}_{C}
\frac {{\rm e}^{- \mu_i r}} {4 \pi r} \equiv K^{(i)}_{C} \:
V_{C} (r,\mu_i)
\label{eq:cpot}
 \\
V_{SS}^{(i)}(r) &=& K^{(i)}_{SS}  \frac{1}{3}
\: \left[ {\mu_i}^2  \: \frac {{\rm e}^{- \mu_i r}} {4 \pi r}
- \delta (r) \right] \equiv K^{(i)}_{SS} V_{SS} (r,\mu_i)
\label{eq:sspot} \\
V^{(i)}_{T} (r) &=& K^{(i)}_{T} \: \frac {1}{3}
\:  \mu_i^2  \: \frac {{\rm e}^{- \mu_i r}} {4 \pi r} \:
\left( 1 + \frac{3} {{\mu_i} r} + \frac{3} {({\mu_i
r})^2} \right)
 \equiv K^{(i)}_{T} V_{T} (r,\mu_i)
\label{eq:tpot}
 \\
V_{PV}^{(i)}(r) &=& K^{(i)}_{PV} \: \mu_i \:
\frac {{\rm e}^{- {\mu_i} r}} {4 \pi r}
\left( 1 + \frac{1} {\mu_i r} \right)
 \equiv  K^{(i)}_{PV} V_{PV}(r,\mu_i)  \ .
\label{eq:pvpot}
\end{eqnarray}
where $\mu_i$ denotes the mass of the
different mesons.
It is these expressions that are inserted in Eq.
(\ref{eq:trel2}) to compute $t_{\rm rel}$ numerically.
The expressions for $K^{(i)}_\alpha$, which contain factors and
coupling constants, are given in Table
\ref{tab:k}.

A monopole form factor $F_{i}({\bf q}^2)=
(\Lambda_i^2-\mu_i^2)/(\Lambda_i^2+{\bf q} ^2)$
is used at each vertex, where the value of the cut-off,
$\Lambda_i$, depends on the meson. We take the values
of the J\"ulich YN interaction\cite{juelich}, displayed
in Table \ref{tab:const} of Section V,
since the Nijmegen model distinguishes form
factors only in terms of the transition channel.
The use of form factors leads to the following
regularization for each meson
\begin{eqnarray}
V_{C} (r; \mu_i) &\to& V_{C} (r; \mu_i) - V_{C}
(r;\Lambda_i) -
\Lambda_i \frac{ {\Lambda_i}^2 - {\mu_i}^2}{2} \frac{{\rm e}^{-
\Lambda_i r}}{4 \pi}
\left( 1 - \frac{2}{\Lambda_i r} \right)  \\
V_{SS} (r; \mu_i) &\to& V_{SS} (r; \mu_i) - V_{SS}
(r;\Lambda_i) -
\Lambda_i \frac{ {\Lambda_i}^2 - {\mu_i}^2}{2} \frac{{\rm e}^{-
\Lambda_i r}}{4 \pi}
\left( 1 - \frac{2}{\Lambda_i r} \right)  \\
V_{ T} (r; \mu_i) &\to& V_{ T} (r; \mu_i) - V_{T}
(r; \Lambda_i) -
\Lambda_i \frac{ {\Lambda_i}^2 - {\mu_i}^2}{2} \frac{{\rm e}^{-
\Lambda_i r}}{4 \pi}
\left( 1 + \frac{1}{\Lambda_i r} \right)  \\
V_{PV} (r; \mu_i) &\to& V_{PV} (r; \mu_i) - V_{PV}
(r; \Lambda_i) -
\frac{ {\Lambda_i}^2 - {\mu_i}^2}{2} \frac{{\rm e}^{- \Lambda_i r}}
{4\pi}
\end{eqnarray}
where $V_{\alpha} (r; \Lambda_i)$ has the same structure as
$V_{\alpha}(r;\mu_i)$, defined in Eqs.
(\ref{eq:cpot})--(\ref{eq:pvpot}), but replacing the meson mass
$\mu_i$ by the corresponding cutoff mass $\Lambda_i$.

\section{The Weak Coupling Constants}

The starting point for describing the weak decay of strange particles
has been the fundamental Cabbibo Hamiltonian based on the Current
$\otimes$ Current assumption
\begin{equation}
H=\frac{G_F}{\sqrt{2}} \int d^{3} x J^{\alpha} (x)
J_{\alpha}^{\dagger}(x) + {\rm h.c.}
\end{equation}
with
\begin{eqnarray}
J_{\alpha} (x) &=& \overline{\Psi}_e(x) \gamma_\alpha (1-\gamma_5)
\Psi_{\nu_e}(x) + \overline{\Psi}_\mu(x) \gamma_\alpha (1-\gamma_5)
\Psi_{\nu_\mu}(x)  \nonumber \\
&+& \overline u (x) \gamma_\alpha (1-\gamma_5) \left( d(x)
\cos{\theta_C} + s(x)\sin{\theta_C} \right) \ ,
\end{eqnarray}
where $\theta_C$ is the Cabbibo angle, $G_F$ the weak coupling
constant and we take the Bjorken and Drell convention for the
definition of $\gamma_5$\cite{bjorken}. As is well known, terms
proportional to $\cos\theta_C$
describe, for instance, the neutron $\beta$-decay while the
contributions
proportional to $\sin\theta_C$ lead to the semileptonic decay of
hyperons and kaons. The $\Delta S$=1 nonleptonic decays are governed by
terms proportional to $\sin\theta_C\cos\theta_C$ which consist of
products of a current between $u$ and $d$ quarks ($\Delta I$=1) and a
current betwen $u$ and $s$ quarks ($\Delta I$ = 1/2). Thus, since terms
in $\sin\theta_C\cos\theta_C$ describe transitions with $\Delta I$ =
1/2 and 3/2 with equal probability, the empirical $\Delta I$ = 1/2 rule
indicates the presence of some dynamical effect related to QCD
corrections that suppresses the $\Delta I$=3/2 components of the
Hamiltonian.

In order to obtain hadronic weak matrix elements of the kind $\langle
M B' | H_w | B \rangle$, where M can stand for pseudoscalar or
vector mesons and B for baryons, it has
been convenient to express the effective weak Hamiltonian in terms of
the SU(6)$_w$ symmetry that unites them.

The $\Delta$S=1 weak nonleptonic Hamiltonian can be written in SU(3) tensor
notation:
\begin{equation}
H_w = \frac{G_F}{2 \sqrt{2}} \cos\theta_C\sin\theta_C \{ J^{2}_{\mu
1}, J^{\mu 1}_{3} \} + {\rm h.c.}
\end{equation}
where
$J^{i}_{\mu j} = (V_{\mu}-A_{\mu})^{i}_{j}$
is the weak hadronic current with SU(3) indices $i$ and $j$. As shown
in Ref.\cite{holstein,delatorre,bala}
the weak vector and axial currents can be
 expressed in terms of SU(6$)_w$ currents.
Since the Hamiltonian is the product of two currents, each belonging to the
$\bf 35$
representation, one can expand
\begin{equation}
35 \otimes 35 = 1_s \oplus 35_s
\oplus 189_s \oplus 405_s \oplus 35_a \oplus 280_a \oplus
\overline{280_a} \ ,
\end{equation}
which allows extraction of the parity violating (PV) and parity
conserving (PC) pieces of the Hamiltonian:
\begin{equation}
H_{\rm\scriptscriptstyle PC}: 1_s \oplus 35_s \oplus 189_s \oplus 405_s
\end{equation}
\begin{equation}
H_{\rm\scriptscriptstyle PV}: 35_a \oplus 280_a \oplus \overline{280_a} \ .
\end{equation}
Each of the possible ways of coupling baryons to mesons within
the SU(6$)_w$ symmetry introduces a reduced matrix element that can
either be fitted to experimental data or calculated microscopically from
quark models. Below we discuss the PV and PC amplitudes separately.

\subsection{ The PV amplitudes}

The traditional approximation employed to obtain the PV amplitudes for
the nonleptonic decays $B \rightarrow B' + M$ has been the use of the
soft-meson reduction theorem:

\begin{equation}
\lim_{q \to 0} \langle B' M_i (q) | H_{\rm pv} | B \rangle
=- \frac{i}{F_{\pi}} \langle B'|[F^5_{i},H_{\rm pv}]|B \rangle
= -\frac{i}{F_{\pi}} \langle B'|[F_{i}, H_{\rm pc}]|B \rangle
\label{eq:soft}
\end{equation}
where q is the momentum of the meson and $F_i$ is an SU(3) generator
whose action on a baryon $B_j$ gives:
\begin{equation}
F_i |B_j \rangle = i f_{ijk} |B_k \rangle \ .
\end{equation}
Since the weak Hamiltonian $H_w$ is assumed to transform like the sixth
component of an octet, a term like $\langle B_k| H^{6}_{w} | B_j \rangle$ can be
expressed as:
\begin{equation}
\langle B_k| H^{6}_{w} | B_j \rangle = i F f_{6jk} + D d_{6jk}
\end{equation}
where $f_{ijk}$ and $d_{ijk}$ are the SU(3) coefficients and $F$ and
$D$ the reduced matrix elements.

With the use of these soft-meson techniques and the SU(3) symmetry
one
can now relate the physical amplitudes of the nonleptonic hyperon decays
into a pion plus a nucleon or a hyperon, $B \rightarrow B' + \pi$, with
the unphysical amplitudes of the other members of the meson octet, the
kaon and the eta. One obtains relations such as\cite{delatorre}:
\begin{eqnarray}
\langle  n K^0 | H_{\rm pv} | n  \rangle &=&
\sqrt{\frac{3}{2}} \Lambda^0_{-} - \frac{1}{\sqrt{2}} \Sigma^+_0
\nonumber \\
\langle  p K^0 | H_{\rm pv} | p  \rangle &=&
-\sqrt{2} \Sigma^+_0
\nonumber \\
\langle  n K^+ | H_{\rm pv} | p  \rangle &=&
\sqrt{\frac{3}{2}} \Lambda^0_{-} + \frac{1}{\sqrt{2}} \Sigma^+_0
\nonumber \\
\langle  n \eta | H_{\rm pv} | \Lambda  \rangle &=&
\sqrt{\frac{3}{2}} \Lambda^0_{-}
\end{eqnarray}
where $\Sigma^{+}_{0}$ stands for $\langle p \pi^{0} | H_{\rm pv}| \Sigma^{+}
 \rangle$,
the PV amplitude of the decay $\Sigma^+ \rightarrow p \pi^0$, which is
experimentally accessible.
We have used the standard notation according to which the hyperon and
meson charges appear as superscript and subscript, respectively.
We point out that using the isospin structure of the potential defined
in the previous chapter,  the  $\langle  N K | H_{\rm pv} | N  \rangle$ matrix
elements are connected to the coupling constants
$C^{\rm\scriptscriptstyle PV}_{\rm\scriptscriptstyle K}$ and
 $D^{\rm\scriptscriptstyle PV}_{\rm\scriptscriptstyle K}$ of Eq.
(\ref{eq:kaweak}) via
\begin{eqnarray}
\langle  n K^0 | H_{\rm pv} | n  \rangle &=&
C^{\rm\scriptscriptstyle PV}_{\rm\scriptscriptstyle K} +
D^{\rm\scriptscriptstyle PV}_{\rm\scriptscriptstyle K}  \nonumber \\
\langle  p K^0 | H_{\rm pv} | p  \rangle &=&
D^{\rm\scriptscriptstyle PV}_{\rm\scriptscriptstyle K}  \nonumber \\
\langle  n K^+ | H_{\rm pv} | p  \rangle &=&
C^{\rm\scriptscriptstyle PV}_{\rm\scriptscriptstyle K}
\end{eqnarray}
As shown above, the symmetry of SU(3) allows connecting the
amplitudes of
the physical pionic decays with those of the unphysical decays involving
etas and kaons. SU(6$)_w$, on the other hand, furthermore permits
relating the amplitudes involving pseudoscalar mesons with those of the
vector mesons. For details we refer the reader to
Refs.\cite{holstein,delatorre};
we just list here the final relations in terms of the coupling
constants defined in the previous section, rather than matrix elements
\begin{eqnarray}
A_\pi&=&\frac{1}{\sqrt{2}} \Lambda^{0}_{-}
\nonumber \\
A_\eta&=&\sqrt{\frac{3}{2}} \Lambda^{0}_{-}
\nonumber \\
C^{\rm\scriptscriptstyle PV}_{\rm\scriptscriptstyle K} &=&
\sqrt{\frac{3}{2}} \Lambda^0_{-} + \frac{1}{\sqrt{2}} \Sigma^+_0
\nonumber \\
D^{\rm\scriptscriptstyle PV}_{\rm\scriptscriptstyle K} &=&
-\sqrt{2} \Sigma^+_0
\nonumber \\
A_{\rho} &=& \varepsilon_{\rho} =
\frac{2}{3} \Lambda^0_{-} - \frac{1}{\sqrt{3}} \Sigma^+_0  + \sqrt{3}
a_T
\nonumber \\
A_{\omega} &=& \varepsilon_{\omega} =
\Sigma^+_0  - \frac{1}{3} a_T
\nonumber \\
C^{\rm {\scriptscriptstyle PV}}_{\rm {\scriptscriptstyle K^*}} &=&
- \sqrt{3} \Lambda^0_{-} + \frac{1}{3} \Sigma^+_0
+ \frac{10}{3} a_T
\nonumber \\
D^{\rm\scriptscriptstyle PV}_{\rm\scriptscriptstyle K^*} &=&
-\frac{2}{3} \Sigma^+_0
+ \frac{8}{9} a_T
\end{eqnarray}
The numerical values of the constants are given in Table
\ref{tab:const}.
Note, that an additional parameter, $a_T$, is present in the coupling
constants for the vector mesons. This coupling, which is very small in
the case of pion emission due to PCAC, can be calculated in the
factorization approximation where the vector meson is coupled to the
vacuum by one of the weak currents. We use the numerical value of
$a_T=-0.953\times 10^{-7}$ from Ref.\cite{delatorre}.

\subsection{The parity-conserving amplitudes}

A description of the physical nonleptonic decay amplitudes $B \rightarrow
B' + \pi$ can also be performed by using a lowest-order chiral analysis.
Employing a chiral lagrangian truncated at lowest order in the energy
expansion for the PV (or s-wave) amplitudes yields results identical to
those discussed above for pseudoscalar mesons. However if one defines
the lowest-order chiral lagrangian for PC (or p-wave) amplitudes, one
finds that such an operator has to vanish since it has the wrong
transformation properties under CP. Thus, the only allowed chiral
lagrangian at lowest order can generate PV but not PC terms.

The standard method to compute the PC amplitudes is the so-called pole
model. As shown in Ref.\cite{donoghue}, this approach can be motivated
by considering the transition amplitude for the nonleptonic emission of
a meson
\begin{equation}
\langle  B' M_i (q) | H_w | B \rangle =
\int d^4x {\rm e}^{i q x} \theta(x^0) \langle B' |
[ \partial A_i(x),H_w(0)] | B \rangle \ .
\end{equation}
Inserting a complete set of intermediate states, $\{|n\rangle\}$,
one can show that
\begin{equation}
\langle  B' M_i (q) | H_w | B \rangle =
-\int d^3x {\rm e}^{i q x} \langle B' |
[ A^0_i(x,0),H_w(0)] | B \rangle - q_\mu M_i^\mu \ ,
\label{eq:pole1}
\end{equation}
where
\begin{eqnarray}
M_i^\mu &=& (2\pi)^3 \sum_n \left[ \delta({\bf p}_n - {\bf p}_{B'} -
{\bf q} ) \frac{\langle B' | A^\mu_i(0) | n \rangle \langle n |
H_w(0) | B \rangle}{p_B^0 - p_n^0} \right. \nonumber \\
&+& \left. \delta({\bf p}_B - {\bf p}_{n} -
{\bf q} ) \frac{\langle B' | H_w(0) | n \rangle \langle n |
A^\mu_i(0) | B \rangle }{p_B^0 - q^0 - p_n^0} \right] \ .
\label{eq:pole2}
\end{eqnarray}

While the first term in Eq. (\ref{eq:pole1}) becomes the commutator
introduced in Eq. (\ref{eq:soft}),
the second term contains contributions from the $\frac{1}{2}^+$ ground
state baryons which are singular in the SU(3) soft meson limit. These
pole terms become the leading contribution to the PC amplitudes. We note
in passing that in principle, such baryon pole terms can also contribute
to the PV amplitudes, however, more detailed studies\cite{donoghue}
showed that their
magnitude is only several per cent of the leading current algebra
contribution.

\begin{figure}
       \setlength{\unitlength}{1mm}
       \begin{picture}(100,180)
      \put(25,0){\epsfxsize=12cm \epsfbox{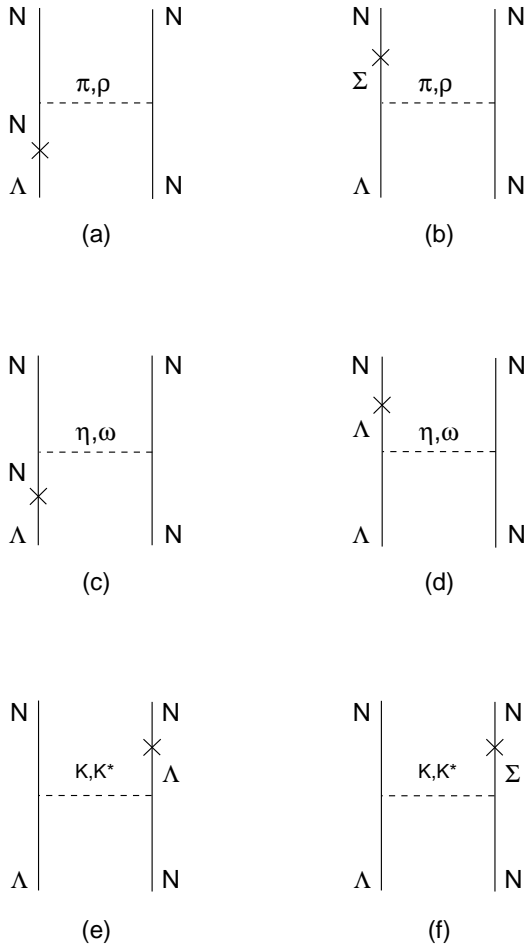}}
       \end{picture}
\caption{
Baryon pole diagrams contributing to the PC weak vertices in the
$\Lambda N \to N N$ transition amplitude.
}
\label{fig:barpole}
\end{figure}

We begin by computing the p-wave amplitude of the
$\Lambda \rightarrow N \pi$ decay since here we can compare with
experiment. The contribution to the PC weak vertex coming from
the baryon pole diagrams shown in Figs. \ref{fig:barpole}(a) and (b)
are given by

\begin{equation}
B_{\pi} =
 g_{\rm{\scriptscriptstyle NN}\pi}\frac{1}{m_\Lambda-m_N}A_{N\Lambda}
+ g_{\rm{\scriptscriptstyle \Lambda \Sigma} \pi}
\frac{1}{m_N-m_\Sigma}A_{N\Sigma}
\end{equation}
where $A_{N \Lambda}$ and $A_{N \Sigma}$ are weak baryon $\rightarrow$
baryon transition amplitudes.
These quantities can be determined via current algebra/PCAC as before
\begin{eqnarray}
\lim_{q\rightarrow 0}\langle \pi^0n|H_{\rm pv}|\Lambda\rangle &=&
{-i\over F_\pi}\langle n|[F^5_{\pi^0},H_{\rm pv}]|\Lambda\rangle =
{i\over 2F_\pi}\langle n| H_{\rm pc}|\Lambda\rangle\nonumber\\
\lim_{q\rightarrow 0}\langle \pi^0p| H_{\rm pv}|\Sigma^+\rangle&=&
{-i\over F_\pi}\langle p|[F^5_{\pi^0}, H_{\rm pv}]|\Sigma^+\rangle=
{i\over 2F_\pi}\langle p| H_{\rm pc}|\Sigma^+\rangle \ .
\end{eqnarray}
Then assuming no momentum dependence for the baryon S-wave decay
amplitude and absorbing the $i$ factor in the definitions of
$A_{N\Lambda}$ and $A_{N\Sigma}$, we obtain
\begin{eqnarray}
A_{N\Lambda} &=&
i\langle n| H_{\rm pc}|\Lambda\rangle =
2 F_\pi\langle\pi^0n| H_{\rm pv}|\Lambda\rangle =
- \sqrt{2} F_\pi\langle\pi^- p| H_{\rm pv}|\Lambda\rangle
= - 4.32\times 10^{-5}\,\,\mbox{MeV}\nonumber\\
A_{N\Sigma} &=&
\frac{i}{\sqrt{2}} \langle p| H_{\rm pc}|\Sigma^+\rangle =
\sqrt{2}F_\pi\langle \pi^0p| H_{\rm pv}|\Sigma^+\rangle
= - 4.35\times 10^{-5}\,\,\mbox{MeV}.
\end{eqnarray}

For the $\eta$ contribution the PC $\Lambda {\rm N} \eta$ term, shown
in Figs. \ref{fig:barpole}(c) and (d),  can be written as:
\begin{equation}
B_{\eta} =
g_{\rm {\scriptscriptstyle NN} \eta}\frac{1}{m_\Lambda-m_N}A_{N\Lambda}
+ g_{\rm {\scriptscriptstyle \Lambda\Lambda} \eta}
\frac{1}{m_N-m_\Lambda}A_{N\Lambda}
\end{equation}
while for the kaon (Figs. \ref{fig:barpole}(e) and (f)), the
expressions are:
\begin{eqnarray}
\langle  n K^+ | H_{\rm pc} | p  \rangle =
C^{\rm\scriptscriptstyle PC}_{\rm\scriptscriptstyle K} &=&
 g_{\rm\scriptscriptstyle{\Lambda p K^+}}\frac{1}{m_n-m_\Lambda}A_{n\Lambda}
+ g_{\rm\scriptscriptstyle{p\Sigma^0 K^+}}
\frac{1}{m_n-m_\Sigma^0}A_{n\Sigma^0}  \nonumber \\
&=& g_{\rm\scriptscriptstyle{\Lambda N K}}\frac{1}{m_N-m_\Lambda}A_{N\Lambda}
- g_{\rm\scriptscriptstyle{N\Sigma K}}
\frac{1}{m_N-m_\Sigma}A_{N\Sigma} \\
\langle  p K^0 | H_{\rm pc} | p  \rangle =
D^{\rm\scriptscriptstyle PC}_{\rm\scriptscriptstyle K} &=&
 g_{\rm\scriptscriptstyle{p\Sigma^+ K^0}}
\frac{1}{m_p-m_\Sigma^+}A_{p\Sigma^+}  \nonumber \\
&=& 2 g_{\rm\scriptscriptstyle{N\Sigma K}}
\frac{1}{m_N-m_\Sigma}A_{N\Sigma} \ ,
\end{eqnarray}
where we have used the relations $A_{n\Sigma^0}=-A_{N\Sigma}$ and
$A_{p\Sigma^+}=\sqrt{2}A_{N\Sigma}$.

The expressions for the vector mesons are similar:

\begin{equation}
\alpha_{\rho} =
 g^{\rm \scriptscriptstyle V}_{\rm {\scriptscriptstyle NN}
 \rho}\frac{1}{m_\Lambda-m_N}A_{N\Lambda}
+ g^{\rm\scriptscriptstyle V}_{\rm {\scriptscriptstyle \Lambda\Sigma} \rho}
\frac{1}{m_N-m_\Sigma}A_{N\Sigma}
\end{equation}
\begin{equation}
\alpha_{\omega} =
g^{\rm\scriptscriptstyle V}_{\rm {\scriptscriptstyle NN}
 \omega}\frac{1}{m_\Lambda-m_N}A_{N\Lambda}
+ g^{\rm\scriptscriptstyle V}_{\rm {\scriptscriptstyle \Lambda\Lambda} \omega}
\frac{1}{m_N-m_\Lambda}A_{N\Lambda}
\end{equation}
\begin{equation}
C^{\rm\scriptscriptstyle PC,V}_{\rm\scriptscriptstyle K^*} =
g^{\rm\scriptscriptstyle V}_{\rm\scriptscriptstyle{\Lambda N K^*}}
\frac{1}{m_N-m_\Lambda}A_{N\Lambda}
- g^{\rm\scriptscriptstyle V}_{\rm\scriptscriptstyle{N\Sigma K^*}}
\frac{1}{m_N-m_\Sigma}A_{N\Sigma}
\end{equation}
\begin{equation}
D^{\rm\scriptscriptstyle PC,V}_{\rm\scriptscriptstyle K^*} =
2 g^{\rm\scriptscriptstyle V}_{\rm\scriptscriptstyle{N\Sigma K^*}}
\frac{1}{m_N-m_\Sigma}A_{N\Sigma} \ ,
\end{equation}
and the tensor coupling constants $\beta_\rho$,
$\beta_\omega$,
$C^{\rm\scriptscriptstyle PC,T}_{\rm\scriptscriptstyle K^*}$ and
$D^{\rm\scriptscriptstyle PC,T}_{\rm\scriptscriptstyle K^*}$
are obtained from the previous expressions by replacing the strong
vector couplings with the tensor ones. The numerical values of all these
coupling constants can be found in Table \ref{tab:const}.

\begin{figure}
       \setlength{\unitlength}{1mm}
       \begin{picture}(100,180)
      \put(25,0){\epsfxsize=12cm \epsfbox{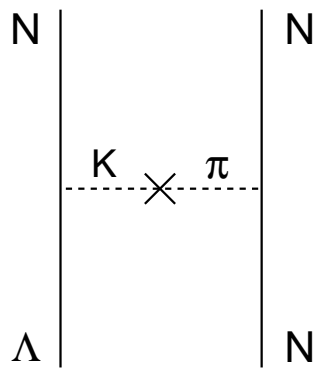}}
       \end{picture}
\caption{
Meson pole diagram contributing to the $\Lambda N \to N N$
transition amplitude.
}
\label{fig:mespole}
\end{figure}

We point out that some studies have included meson pole diagrams of the
form shown in Fig. \ref{fig:mespole}. The contribution of these
diagrams would be given by
\begin{equation}
g_{\rm\scriptscriptstyle \Lambda NK}\frac{1}{m_{\rm \scriptscriptstyle
 K}^2-m_\pi^2}
A_{\rm {\scriptscriptstyle K} \pi}
\end{equation}
where the meson $\rightarrow$ meson weak transition amplitude
$A_{\rm {\scriptscriptstyle K} \pi}$ can also be related via PCAC to the
experimental amplitude
for $K \rightarrow \pi \pi$ decay, yielding
$A_{\rm {\scriptscriptstyle K} \pi} = - 2.5 \times
10^{-3}$ MeV$^2$\cite{donoghue}. There is considerable uncertainty
regarding the phase between the meson and the baryon poles which lead
some studies to adjust it to better reproduce the data.
It has been argued\cite{donoghue}
 that the presence of
these meson pole diagrams is important to fulfil the requirements of the
so-called Feinberg-Kabir-Weinberg theorem in the nonleptonic decays. On
the other hand, counting powers of energy in a chiral analysis, one
finds that while the baryon pole terms are of order $q^{-1}$ the meson
poles enter at next order, along with higher-order chiral lagrangians.
In general, we found these contributions to be very small and have
therefore neglected them in the following. In principle, SU(6$)_w$ can
be used as well to relate the weak meson $\rightarrow$ meson
pseudoscalar transition amplitudes with those of the vector mesons. The
results, given here for completeness, are:
\begin{eqnarray}
A_{\rm{\scriptscriptstyle K} \eta} &=& - \frac{1}{\sqrt{3}} A_{\rm
 {\scriptscriptstyle K} \pi}
\nonumber \\
A_{\rm{\scriptscriptstyle K^*} \rho} &=&  A_{\rm{\scriptscriptstyle K} \pi}
\nonumber \\
A_{\rm{\scriptscriptstyle K^*} \omega} &=& - \frac{1}{\sqrt{3}}
 A_{\rm{\scriptscriptstyle K} \pi}
\ .
\end{eqnarray}

\section{Results}
\subsection{$\pi$-Exchange}

We begin our discussion by presenting the results using
only the OPE part of the
weak $\Lambda N \to NN$ interaction. On one hand
we would expect this meson to adequately describe at the least the
long-range
part of the transition potential, while on the other hand its contribution
has minimal uncertainties since the weak $\Lambda {\rm N} \pi$ vertex
is experimentally known.
It is therefore a good starting point to assess the significance of form
factors as well as initial and final state correlations before including the
other mesons in the potential.  The results of our calculations with OPE only
are shown in Table \ref{tab:pi} where the
nonmesonic decay rate of $^{12}_\Lambda$C is given in units of the free lambda
decay rate ($\Gamma_\Lambda$). The uncorrelated results (FREE) are
compared with computations that include
initial $\Lambda$N short-range correlations (SRC),
form factors (FF), and final-state
interactions (FSI) separately for the central (C), tensor (T)
(adding to a total parity-conserving (PC) contribution) and
parity-violating (PV) potentials.  The free central term is reduced
dramatically by the initial SRC,
however, most of the uncorrelated central potential contribution is in
fact due to the $\delta$-function in the transition potential
which is completely eliminated by SRC.
Without the $\delta$-function, the central part is reduced by about a
factor of two.
Including SRC, FF, and FSI gives a negligible central
decay rate. In contrast, the contribution of the tensor interaction
is reduced only 10\% by SRC and by 20-35\% once FF and FSI
are included as well.  Therefore, the contribution of the central term
amounts to less than 0.5\% of the
total $\pi$-exchange rate.  This behavior has been found and discussed
by other authors as well\cite{mckellar,dubach,takeuchi}.
On the other hand, our
PV potential yields about 40\% of the $\pi$-exchange rate, at variance
with older nuclear matter results that reported either a
15\%\cite{dubach} or a negligible\cite{mckellar} PV contribution
to the rate.

\begin{figure}
       \setlength{\unitlength}{1mm}
       \begin{picture}(100,180)
      \put(25,0){\epsfxsize=12cm \epsfbox{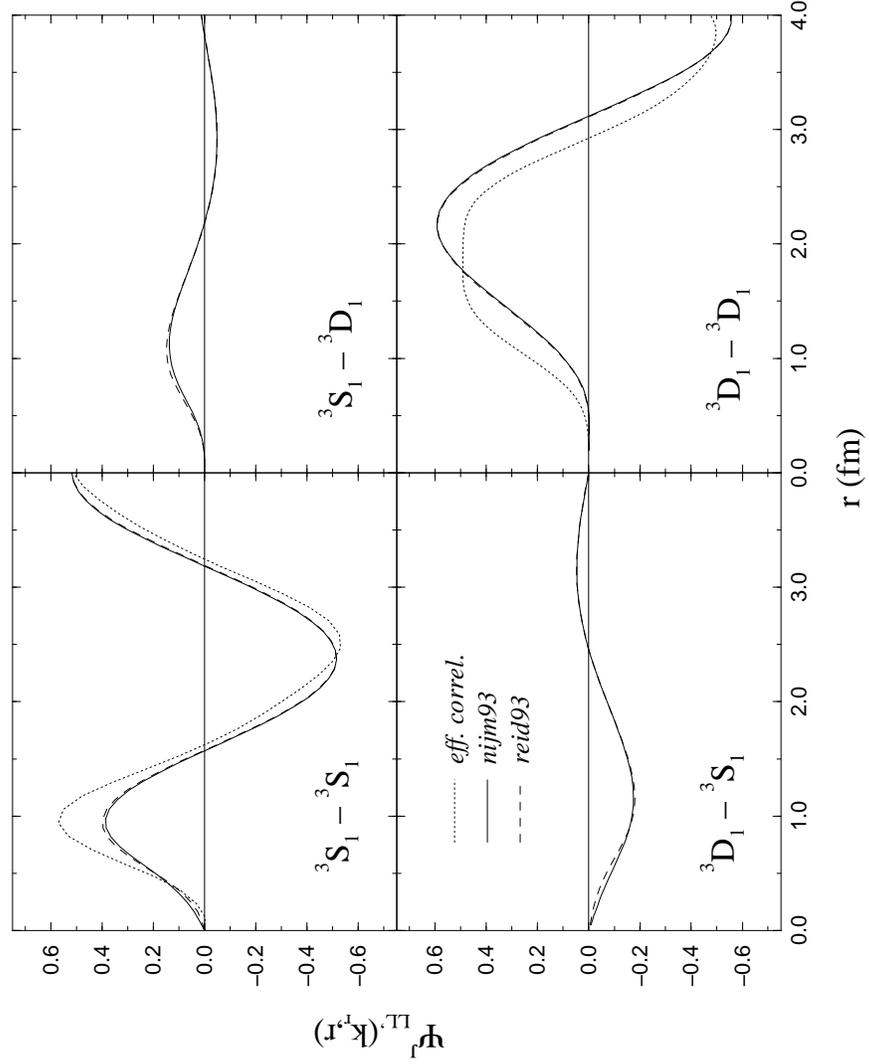}}
       \end{picture}
\caption{
$^3S_1-^3D_1$ coupled channel NN wave functions for a relative
momentum of $k_r=1.97$ fm$^{-1}$, obtained with the Nijmegen93 (solid
line) and the Reid93 (dashed line) interactions. The dotted line
represents
the phenomenological correlated wave function discussed in the text.
}
\label{fig:coup}
\end{figure}

\begin{figure}
       \setlength{\unitlength}{1mm}
       \begin{picture}(100,180)
      \put(25,0){\epsfxsize=12cm \epsfbox{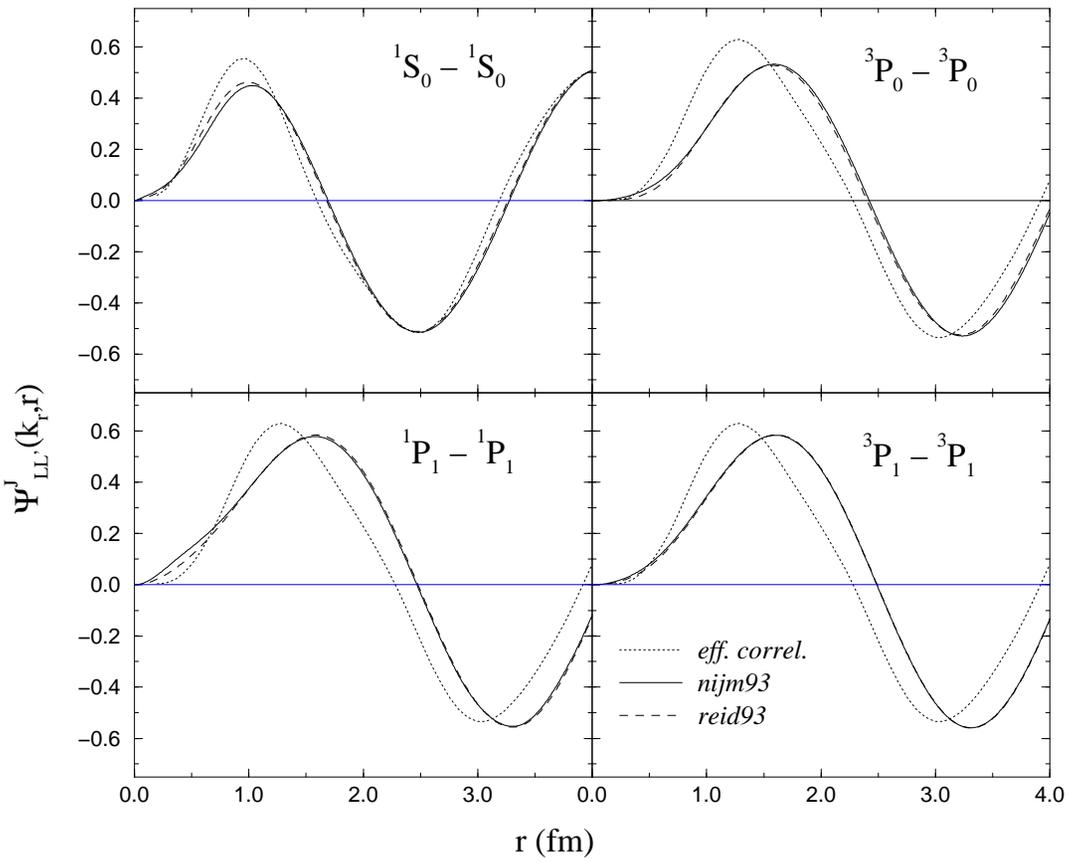}}
       \end{picture}
\caption{
Same as Fig. 5 for the uncoupled $^1S_0$, $^3P_0$, $^1P_1$ and
$^3P_1$ channels.
}
\label{fig:nocoup}
\end{figure}

In previous papers\cite{sitges,assum} we have demonstrated the
sensitivity of the calculated decay rates to the form of the initial SRC. In
particular, we found that older calculations using a phenomenological
NN
correlation function\cite{mckellar,kisslinger} to simulate $\Lambda$N
SRC in the initial state, rather
than SRC based on a realistic YN meson-exchange potential as is done
here, tend to overpredict the amount of initial correlations.
We find similar results for the final-state interactions.
As shown in Table \ref{tab:pi}, the PV and PC rates
are reduced
by about 35\% and 50\%, respectively, when FSI are included via
a correlation function based on a
realistic NN potential (last two columns), rather than the 20\% and
40\% reduction obtained with the phenomenological
NN SRC function of Eq. (\ref{eq:cor}).
It is comforting to see that the variation between different
realistic NN interactions,
such as the soft-core Nijmegen and a modern version of the Reid
potential, plays essentially no role.
This behavior can also be understood from Figs. \ref{fig:coup} and
\ref{fig:nocoup}, where we compare the
different correlated wave functions for several channels and a
relative momentum of $k_r=1.97$ fm$^{-1}$.
The figures demonstrate that the phenomenological
correlated wave function overestimates
the realistic NN wave functions at intermediate distances ($0.5-1.5$
fm) while it underestimates them at short distances. As a result, the
phenomenological approach overestimates the decay rate by about
20\%.
Including realistic final state correlations leads to an interference
between central and tensor transitions. For phenomenological FSI the
central and tensor contributions to the total rate can be seen in
Table \ref{tab:pi} to add incoherently. Once a realistic NN
potential is used this
incoherence is replaced by destructive interference. While this is
only a small effect for the pion due to the small size of
the central potential term this interference is more significant for the
vector mesons where the central transition amplitude is large compared
to the tensor term.

The second quantity of interest displayed in Table \ref{tab:pi} which
is sensitive to
the isospin structure of the transition amplitude is the neutron- to proton-
induced ratio $\Gamma_n/\Gamma_p$. Noticeable is the smallness of the ratio
which is due to the Pauli Principle that suppresses the final $T=1$,
$L=2$, $S=1$ state with its antisymmetrization factor $(1-)^{L+S+T}$.
That excludes the tensor transition in the neutron-induced
rate, which gives rise to $nn$ $(T=1)$ pairs, and it is precisely this
tensor piece which constitutes the largest part of the OPE
diagram.
However, this argument holds only for relative
$\Lambda$N S-states. For nucleons in the p-shell there exits a
relative
$\Lambda$N P-state which contributes a small but nonzero amount of
the tensor potential to the neutron-induced decay.
Note that including initial SRC, FF and realistic
FSI reduces the $\Gamma_n/\Gamma_p$  further by about 40\%. This is
due to the elimination
of the central potential for which we obtain a value for
$\Gamma_n/\Gamma_p$  of about 1/3.
In principle, one would expect $\Gamma_n/\Gamma_p$ =1/2 for the
central term due to the statistical factor of 1/2 that accounts for
two identical particles in the final state.
For $^{12}_\Lambda$C this number becomes 1/2.4 since we have
5 neutrons and 6 protons.
The remaining difference comes from different
$^1S_0$ $(T=1)$ and $^3S_1$ $(T=0)$ final state
wave functions which enter
the various spin-isospin channels and, therefore,
lead to slightly different
 $\Lambda n \to n n$ and $\Lambda p \to n p$ transition
amplitudes.

The suppression of the central potential term due to SRC, FF and FSI also
explains the difference between
the uncorrelated and the fully correlated ratio of PV to PC
amplitudes, PV/PC, shown in Table \ref{tab:pi}
as well.  More relevant than this ratio, however, is the
asymmetry parameter,
$a_{\Lambda}$, defined in Eq. (\ref{eq:asym}).
This quantity, which measures
the interference between the PC and the PV part of the amplitude, can be
accessed
experimentally in contrast to the PV/PC ratio which is merely of theoretical
interest. We find that this asymmetry parameter is only mildly sensitive to
initial SRC and FF but changes by more than a factor of two when realistic
FSI are included. This observable thus clearly demonstrates that for its
accurate prediction the use of a realistic NN potential to describe
the interactions in the final state is imperative.
Below we will use realistic FSI generated with the Nijmegen potential
for all results that include final state correlations.

Comparing our results obtained here with older nuclear matter
computations\cite{salcedo,mckellar,dubach,holstein} we point out that
the fully correlated total rate in nuclear
matter is predicted to be in the range of 1.85 - 2.3, thus overpredicting our
shell model calculations by more than a factor of two.
When a Local Density Approximation (LDA) is
performed\cite{ramos1,salcedo}
the rate reduces to $\Gamma_{nm} = 1.45\, \Gamma_\Lambda$ for
$^{12}_\Lambda$C. Although this value is further reduced when the same
$\Lambda$ wave function as in the present work is used, the LDA result is
still larger by about
20-40\%,
depending on the choice of the Landau-Migdal parameter which
measures the initial state correlations.
The remaining difference may not be
entirely surprising since the LDA is expected to work well for heavier
nuclei and begins to break down for s- and p-shell nuclei.
We also find that our
PV potential yields about 40\% of the $\pi$-exchange rate, at variance
with nuclear matter results that reported negligible PV
rates\cite{mckellar}.

We conclude our discussion of the OPE-only by assessing the role of the
relative $\Lambda$N P-state contributions. In shell model
calculations
such terms naturally arise for nucleons in p-shell and higher orbitals
when one transforms from shell model
coordinates to the relative $\Lambda$N 2-body system. For s-shell
nucleons, where one has an S-wave in both the relative and
center-of-mass (CMS)
motions, the transition amplitude gives a maximum contribution at the
back-to-back kinematics, ${\bf k}_1 = -{\bf k}_2$, yielding a total
CMS momentum ${\bf K}=0$.
For p-shell nucleons one would expect the contribution from the
initial relative $\Lambda$N S-state to be suppressed compared to
that
of the relative P-state since the CMS harmonic oscillator wave function
is then a P-state and thus zero at ${\bf K}=0$. Surprisingly, we find
that after integrating over all kinematics with ${\bf k}_1 \neq
-{\bf k}_2$,
this relative L=0 term contributes about 90~\% to the p-shell
rate\cite{bennhold2}. Thus,
once the whole phase space is included, most of the total decay rate of
p-shell nucleons still comes from the relative $\Lambda$N S-state.
Furthermore, neglecting the relative P-state contribution leaves
the ratio $\Gamma_n/\Gamma_p$
unaltered while the asymmetry parameter
$a_\Lambda$  is reduced by 10\%.

\subsection{$\pi$- and $\rho$-Exchange}
In this section we begin examining the role of additional mesons.
As discussed above, one is faced with the immediate difficulty
that none of the weak couplings involving heavier mesons can be accessed
experimentally. Thus, one is required to resort to models which in this
case involve considerable uncertainty. Table \ref{tab:pirho} presents
our results
for the $\rho$-meson exchange alone as well as for the $\pi$- and
$\rho$-exchanges combined.  Since both the $\Lambda {\rm N} \pi$- and
the $\Lambda {\rm N} \rho$-couplings are
obtained within the same model there is no sign ambiguity.
As noted before,
the central potential can now be divided into a spin-independent
(C) and a spin-dependent (SS) piece, which are shown separately.
In contrast to the
pion case, the factor $m_\rho^2$ in front of the
Yukawa function in the SS central part of Eq. (\ref{eq:sspot})
enhances this contribution which then becomes comparable in
magnitude to the piece containing the delta function. The two terms
interfere
destructively and yield a SS central part that is about half the size
of the tensor contribution. In Ref. \cite{rho} we noted
that SRC reduce
both the central C- and the SS-part of the $\rho$-meson contribution
without $\delta$-function by almost a
factor of 10, compared to a factor of 2
in the $\pi$ case, reflecting the much
shorter range of the $\rho$-exchange diagram.  Similarly, the tensor
interaction of the $\rho$ is reduced by a factor of 2.5, compared
to a 10\% reduction in the $\pi$ case, as soon as SRC are included.
The additional inclusion of FF and FSI further reduces both the central
and tensor rates by substantial amounts.
As is evident from Table \ref{tab:pirho}, the final result of the
central $\rho$ contribution exceeds the $\rho$ tensor term by
almost a factor of two. Due to the strong destructive interference
induced by the FSI between the tensor and the central part the total PC
rate turns out to be smaller than either term alone.

In terms of the combined $\pi$ and $\rho$ contribution
we find destructive interference between the two mesons for the PC rate
but constructive interference for the PV decay mode. While the
$\pi$-only PC rate is reduced by 16\% when the $\rho$ is added, the
$\pi$-only PV rate is enhanced by about 17\% even though the
$\rho$-only
PV rate is very small. These counterbalancing interferences lead
to a combined
$\pi + \rho$ total decay rate that is very similar to that of the
$\pi$ alone. The neutron to proton induced ratio
$\Gamma_n / \Gamma_p$, on the other hand, is slightly decreased.
This could be, in principle, surprising since the tensor rate of the
$\rho$-exchange is not as dominant as it is in the $\pi$ case. As
noted in Ref. \cite{rho}, this is due to an interference pattern of
the central $C$ and $SS$ amplitudes which is destructive for the
n-induced and constructive for the p-induced mechanisms. This
yields a central rate which is basically p-induced.
The strongest change can be seen in the asymmetry parameter $a_\Lambda$
which is reduced by more than a factor of 2.
This reduction can
be traced to the above mentioned interference pattern between the PV and
PC rates which are measured by this observable.

The $\rho$-meson was the first of the heavier meson which was included
in several earlier calculations. The first attempt was due to McKellar
and Gibson\cite{mckellar} in a nuclear matter framework which evaluated
the weak $\Lambda {\rm N} \rho$ couplings using SU(6) and,
alternatively, a factorization
model. In their approach, which neglected the PV couplings,
the phase between the $\pi$ and $\rho$
amplitudes was not determined and their final results varied dramatically
with their different models for the weak coupling constants. In a more
recent calculation\cite{nardulli}, Nardulli obtains the PC couplings in
a pole model approach similar to ours. Besides the ground state baryon
pole he includes the $\frac{1}{2}^*$ baryon resonance pole terms as well
as K$^*$ pole contributions that appear in
Ref. \cite{holstein,delatorre}
but have been omitted here. The weak baryon-baryon transition amplitudes
for the resonance poles are taken from a pole model analysis of hyperon
PC pion decays which uses an F/D ratio of $-1$ for the weak baryon
transition amplitudes and adjusts the overall coupling to the
experimentally measured p-wave $\pi$ decay rates. As pointed out in
Ref. \cite{maltman95} the most serious
problem with this fit is that it employs a $K \rightarrow \pi$ weak
transition amplitude in its K pole graphs that is about an order of
magnitude larger than the strength extracted from the weak kaon decay
mode $K \rightarrow \pi \pi$\cite{donoghue}. The K$^*$ pole
contributions
are calculated using a simplified factorization approach in which a
number of terms are neglected\cite{maltman95}. The PV couplings of
Ref. \cite{nardulli} are computed in a pole model approach that
includes
baryon resonance poles with negative parity, belonging to the (70,$1^-$)
multiplet of SU(6). In order to obtain the weak baryon transition
amplitudes the experimental hyperon s-wave $\pi$ decays are used as
input. Therefore, his approach for the PV weak $\Lambda {\rm N} \rho$
vertex is
considerably different from the analysis used in this study. For the
sake of comparison, Table \ref{tab:pirho} also lists the $\rho$ term
calculated with
Nardulli's weak coupling constants. The PC transition potentials turn
out to be very similar in magnitude while the PV rate is larger by more
than a factor of 10 for the $\rho$ alone. This increase enhances the
$\pi + \rho$ total decay rate by about 25\%, while the
$\Gamma_n/\Gamma_p$ and $a_\Lambda$
are reduced by roughly the same amount. One should point out that the
close
agreement in the PC terms is fortuitous since the baryon resonance
pole terms which are not present in our approach contribute about
30\% to the weak $\Lambda {\rm N} \rho$ tensor coupling of
Ref.\cite{nardulli}. From this comparison it becomes
obvious that there is considerable
uncertainty in the determination of the weak vector meson vertices.

\subsection{K- and K$^*$-Exchange}
The results of our calculations for the exchanges of the strange mesons,
K and K$^*$, are shown in Table \ref{tab:kakst}.
The kaon is the lightest meson after the pion with a strong coupling
constant $g_{\rm {\scriptscriptstyle \Lambda N K}}$ of comparable magnitude to
 $g_{\rm {\scriptscriptstyle N N} \pi}$ but of opposite sign.
The
total decay rate for the K-only exchange diagram amounts to about 15\%
of the $\pi$ term which is the largest contribution among
 all the heavier mesons. As discussed
further below the kaon therefore significantly interferes with the
$\pi$-only rate.

In contrast to the
$\pi$ and the
$\rho$ which are isovector mesons and the $\eta$ and $\omega$ which are
isoscalar, the K and K$^*$ can contribute to both the $T=0$ and
$T=1$ weak
$\Lambda N \to NN$ transition potential and therefore have two independent
 couplings,
$C_{\rm\scriptscriptstyle K}$ and $D_{\rm\scriptscriptstyle K}$ [see Eq.
 (\ref{eq:kaweak})]. Due
to their
isospin structure it was pointed out several years ago\cite{gibson89}
that including the kaon exchange has the potential to strongly influence
the $\Gamma_n/\Gamma_p$ ratio. Using a simple schematic model that
ignores the spin structure it was shown
that the ratio $\Gamma_n/\Gamma_p$  could be estimated with the
expression
\begin{equation}
\frac{\Gamma_n}{\Gamma_p}=\left| \frac{A_0+A_1}{A_0-3A_1}
\right|^2 \ ,
\label{eq:gibson}
\end{equation}
where $A_0$ and $A_1$ are the isoscalar and isovector coupling
constants, respectively, which in our case are given by
\begin{eqnarray}
A_0&=&\frac{C_{\rm\scriptscriptstyle K}}{2}+ D_{\rm\scriptscriptstyle K}
 \nonumber \\
A_1&=&\frac{C_{\rm\scriptscriptstyle K}}{2} \ .
\end{eqnarray}
With the PV values of Table \ref{tab:const}, we obtain
$\Gamma_n/\Gamma_p=4.6$ using Eq. (\ref{eq:gibson}), confirming the
result quoted in Ref.
\cite{gibson89}. However, we obtain
$\Gamma_n/\Gamma_p=0.23$ when the PC coupling constants
of Table \ref{tab:const} are used. Our complete result, which
considers the
spin structure and includes both PC and PV amplitudes, turns out to be
$\Gamma_n/\Gamma_p=0.26$ as shown in Table \ref{tab:kakst}. This much
smaller result suggests that one cannot draw conclusions about
the ability of the strange mesons to drastically increase the neutron
to proton ratio. Moreover, we have calculated this ratio using PV
and PC amplitudes only but retaining the spin dependence and obtain,
respectively, the values
1.69 and
0.03, far away from the estimates made above using
Eq. (\ref{eq:gibson}).

As can be seen in Table \ref{tab:kakst}, the $\Gamma_n/\Gamma_p$
ratio for the kaon only is larger than the corresponding
value for the pion by a factor of 2.5, while the asymmetry parameter
obtained is very small. As discussed in the previous section our
framework for the weak baryon-baryon-meson coupling constants assumes
the validity of SU(3) (and SU(6) in the case of the vector mesons). No
attention has been paid to the effects of SU(3) symmetry breaking which
is known to be of the order of 30\%.
These effects have been addressed in
a recent work by Savage and Springer\cite{spring} in the framework of
next-to-leading order chiral pertubation theory ($\chi$PT).
They point out that
understanding the weak NNK vertex could elucidate a problem
regarding
the nonleptonic $\pi$ decay of free hyperons. While the PV
(s-wave) amplitudes of
these $\Delta S$=1 decays are adequately reproduced at tree level, the
corresponding PC (p-wave) amplitudes cannot be well described using
coupling
constants from the s-waves as input. A one-loop calculation of the
leading SU(3) corrections\cite{jenkins92},
performed in $\chi$PT, found that these loop
corrections can change the tree level prediction of the p-wave
amplitudes by a disturbing 100\%, thus raising questions about the
validity of $\chi$PT in this sector. As an alternative it was
suggested\cite{jenkins92} that large cancellations may occur between
tree-level p-wave $\pi$ decay
amplitudes which would magnify the SU(3) breaking effects. The one-loop
corrections to the weak NNK vertex found in ref.\cite{spring},
 on the other hand, modify the
tree-level p-wave amplitudes by only up to 30\%. If an experimental
signature for these SU(3) corrections could be found in the nonmesonic
decay it would provide insight into the applicability of $\chi$PT to
these reactions. Table \ref{tab:kakst} shows the results of our
calculations
performed with the Savage-Springer weak NNK couplings.
As expected, the kaon rates are roughly a factor of two smaller
since the improved constants are reduced by about 30\%. The values of
the $\Gamma_n/\Gamma_p$ ratio and
the asymmetry, on the other hand,
are barely affected because all
pieces of the transition amplitude are reduced by about the same amount.

The K$^*$ vector meson is the heaviest meson exchanged in our weak
$\Lambda N \to NN$
transition potential. Nevertheless, due to its large weak NNK$^*$ and
strong $\Lambda {\rm N K}^*$ tensor couplings it is more important
than
either the $\rho$ or the $\omega$. The central and tensor potential
contributions are comparable in size, however, due again to the interference
generated by the realistic FSI that mixes S and D states, the total PC
rate turns out to be of the same magnitude.
The larger PV coupling constant yields a PV rate which is significantly
greater than the corresponding rates for the $\rho$ and the $\omega$
exchange contribution. The total K$^*$-only decay rate is seen to be
about half of the K-only rate but twice as large as the $\rho$- and
$\omega$-only
rates. Due to the relative magnitude of the central and PC potential
terms we find a $\Gamma_n/\Gamma_p$  ratio for the K$^*$ diagram which is 0.5,
about five times larger than that for the $\pi$-exchange ratio.
This is a dramatic
illustration of the difference in isospin structure between the various
mesons. The K$^*$ diagram also exhibits more than twice the asymmetry
parameter of the K-meson.

Coincidentally, the ${\rm K}+{\rm K}^*$ total rate is only slightly
larger than the K-only total rate even though the interference
between the K and K$^*$ is important,
as can be seen from the separate channel contributions. The PV rate is
more than twice the K-only result and, as a consequence, the
asymmetry parameter is enhanced dramatically.
The $\Gamma_n/\Gamma_p$ ratio also turns to be quite large,
although it remains to be seen how the
interference between all the mesons actually affects the final result
for the observables.
This is discussed in subsection E.

\subsection{$\eta$- and $\omega$-Exchange}
Table \ref{tab:etaome} presents our results for the isoscalar mesons
alone. The
$\eta$-meson exchange contribution is by far the smallest of the
different mesons included in our potential. This may come as no surprise
since it is known that including the $\eta$ in phase shift fits of NN
potentials influences the parameters only marginally. The main reason
for the small size of the $\eta$-exchange lies in the magnitude of the
strong NN$\eta$ coupling constant, which is not well determined
but is known to be much smaller than both $g_{\rm {\scriptscriptstyle N N} \pi}$
 and $g_{\rm {\scriptscriptstyle \Lambda N K}}$. We have used
the
value of the Nijmegen potential which is $g_{\rm {\scriptscriptstyle N N} \eta}
 = 6.4$ even though the
NN phase shifts are very insensitive to this coupling. Recent data on
$\eta$ photoproduction on the nucleon find a much reduced value of
around $g_{\rm {\scriptscriptstyle N N} \eta} = 1.4$\cite{tiator94}. Should
 these conclusions be
confirmed
then clearly the $\eta$ can safely be neglected in both the strong
NN
sector as well as the weak $\Lambda N \to NN$ transition potential discussed
 here.
As shown in Table \ref{tab:etaome}, the behavior of the $\eta$
contribution follows
that of the $\pi$ term: a negligible central term, the largest piece
coming from the tensor potential, and a PV rate about half the size of
the tensor term. Note that in contrast to the pion there is no charge
exchange term for the proton-induced decay, thus the
$\Gamma_n/\Gamma_p$ for the $\eta$
only is larger than that for the pion by almost a factor of four.

The $\omega$-meson exchange provides an interesting contrast to the
$\rho$ contribution since for the strong NN$\rho$ vertex the
vector coupling is relatively small and the tensor coupling is large,
while the reverse is true for the strong NN$\omega$ couplings. The
weak PC and PV couplings (see Table \ref{tab:const}), on the other
hand,
are comparable in size. This pattern of strong couplings is reflected
in the distribution of decay strength seen in Table \ref{tab:etaome}.
The largest contribution comes from
the spin independent central term which is about twice as large as the
corresponding
term for the $\rho$. The $\omega$ tensor term, however, is smaller than
the $\rho$ tensor potential by about a factor of seven. The interference
of the various terms yield a total PC rate for the $\omega$ meson that
is very similar in
magnitude to the $\rho$ but generates a much larger
$\Gamma_n/\Gamma_p$. Similar to the
$\rho$ the PV rate is negligible due to the small size of the weak PV
$\Lambda {\rm N} \omega$ coupling constant. No other models are
available for weak vertices involving this meson.

\subsection{The full weak one-meson-exchange potential}

\begin{figure}
       \setlength{\unitlength}{1mm}
       \begin{picture}(100,180)
      \put(25,0){\epsfxsize=12cm \epsfbox{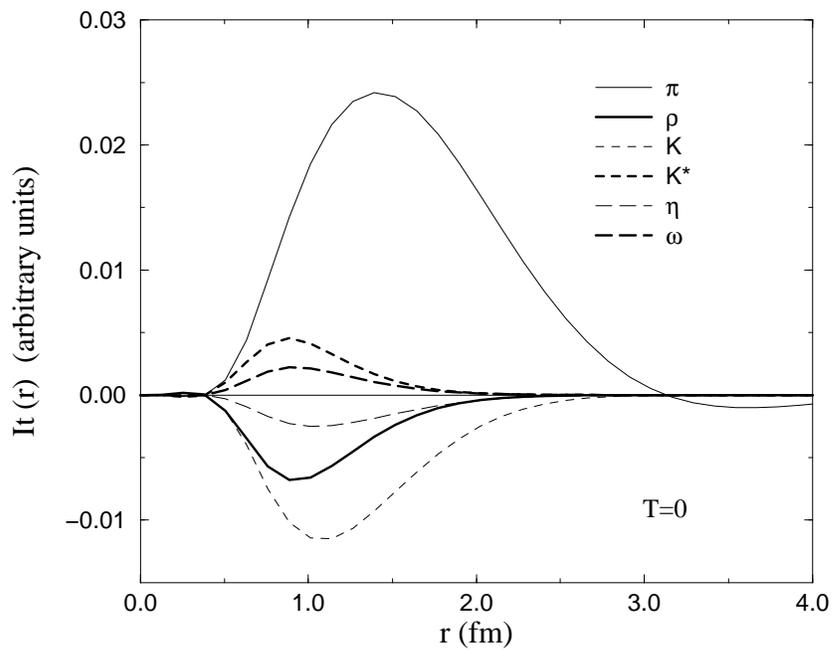}}
       \end{picture}
\caption{
Contribution of the different mesons to the integrand of the
$^3S_1-^3D_1$ ($T=0$) correlated weak transition amplitude.
}
\label{fig:tensor}
\end{figure}

\begin{figure}
       \setlength{\unitlength}{1mm}
       \begin{picture}(100,180)
      \put(25,0){\epsfxsize=12cm \epsfbox{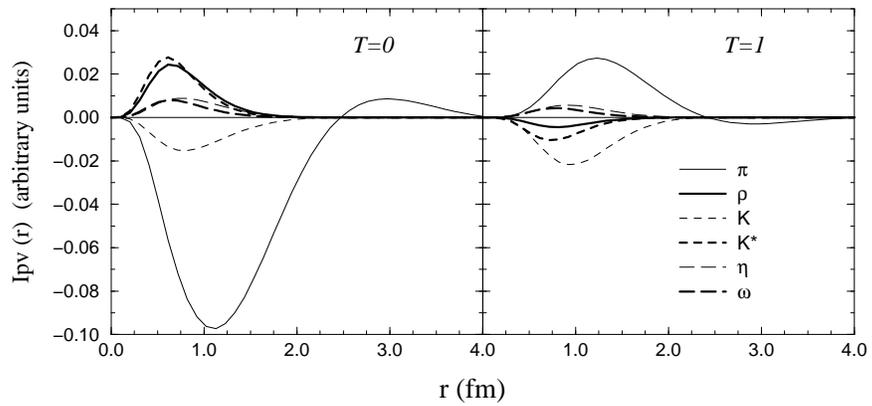}}
       \end{picture}
\caption{
Same as Fig. 7 for the PV $^3S_1-^1P_1$ ($T=0$) and $^3S_1-^3P_1$
($T=1$) transition amplitudes.
}
\label{fig:pv}
\end{figure}

In this subsection, we explore the effect of including
all the mesons discussed before on the weak decay observables. For
this purpose we show in Figs. \ref{fig:tensor} and \ref{fig:pv}
 the contribution of the
different mesons to the integrand of Eq. (\ref{eq:trel2}) for relevant
transition channels.  Fig. \ref{fig:tensor} displays the tensor
transition
$^3S_1 \to {}^3D_1$ $(T=0)$
of the PC amplitudes since it yields the most important
contribution for pseudoscalar mesons and gives rise to
important interference effects when the contributions of the mesons
are added in pairs of identical isospin.
As is evident from the figure, the pion-exchange contribution dominates,
not only in magnitude but also in range;  a consequence
of the pion being the lightest meson.  As expected, the kaon provides the
second-largest contribution with a range somewhat less than that of the
pion, followed by the heavier mesons with an even shorter range. Note
that the contribution of each isospin-like pair
[$(\pi,\rho)$, (K,K$^*$), $(\eta,\omega)$] interferes destructively,
thus the large tensor contribution of each
pseudoscalar meson is partially cancelled by that of its vector
meson partner, an effect that can also be explicitly seen in Tables
\ref{tab:pirho}, \ref{tab:kakst} and \ref{tab:etaome}, discussed above.
We have not shown the integrands of the central transitions since they
are very small for the pseudoscalar mesons.

Significant interferences are also observed for the integrands of
the PV transitions $^3S_1 \to {}^1P_1$
($T=0$) and $^3S_1 \to {}^3P_1$ ($T=1$), shown in Fig. \ref{fig:pv}.
Again, we find the pion to be dominant among the mesons
 in the $T=0$ transition,
while the contribution of the other mesons play a more
important role in the $T=1$ channel.

Below we discuss results for the different
observables characteristic of the weak decay. We compare
the results obtained with the Nijmegen strong coupling constants with those
obtained using the J\"ulich strong couplings given in Table
\ref{tab:const}.
Although in principle the strong couplings also affect the PC weak vertices
through the pole model, our goal here is
to assess, for one particular model of weak couplings, the effect
of using strong coupling constants from two different
YN potentials which fit the hyperon-nucleon scattering data equally
well.

The results in Table \ref{tab:rate} again demonstrate the significance of
the short-range correlations and form factors in the nonmesonic decay.
Adding the heavier mesons without form factors and SRC (column FREE)
leads to a total rate that fluctuates significantly, with
the additional mesons giving an
appreciable contribution to the $\pi$-exchange rate.
This behavior is considerably suppressed by short-range
effects, as shown in the second column.
The rate is especially sensitive to the inclusion of the strange
mesons. While including the $\rho$-meson has almost no effect
the addition of kaon exchange reduces the total rate by almost
50\% when the Nijmegen strong couplings are used.
The reduction is mostly compensated by the addition of the K$^*$,
yielding a rate 15\% below the pion-only decay rate.  The situation is
similar for the $\eta$ and $\omega$, their combined effect on the rate
is negligible.  Thus, with Nijmegen couplings adding the
heavier mesons gives a reduction of only 15\%.  The situation is slightly
different when the J\"ulich strong coupling constants are employed;
their omission of the $\eta$ and their larger K$^*$ and $\omega$
couplings lead to a total rate 15\% larger than the pion-only rate.
This indicates that the results are sensitive to
the model used for the strong vertices, although both results are
consistent with the present experimental values.
This sensitivity to the strong coupling constants is unfortunate since
it will certainly complicate the task of extracting weak couplings from
this reaction. Improved YN potentials which narrow the range of the
strong coupling constants are required to reduce this uncertainty.
Table \ref{tab:rate} also shows the results obtained when the NNK
weak coupling constants derived in next-to-leading order
$\chi$PT\cite{spring} are used. Due to the smaller value of the
coupling constants the effect of the K meson is reduced and thus
the total rate is increased by about 10\%.

The results for the ratio of the neutron- to proton-induced partial
rates $\Gamma_n/\Gamma_p$ are shown in Table \ref{tab:weakobs}.
The neutron- to proton-induced ratio is, as
expected, quite sensitive to the isospin structure of the exchanged
mesons. It has been known for a long time that pion exchange alone
produces only a small ratio\cite{mckellar}.
While the role of the $\rho$ is limited
it is again the inclusion of the two strange
mesons that dramatically modifies this partial ratio.
Including the K-exchange which interferes
destructively with the pion amplitude in the neutron-induced channel
(see, for example, the $T=1$ PV transition amplitudes of Fig.
\ref{fig:pv}) leads to a reduction
of the ratio by more than a factor of three.
The K$^*$, on the other hand, adds contructively.
Again, an indication of this behavior can be seen in
Figs. \ref{fig:tensor} and
\ref{fig:pv}. In the $T=1$ PV channel, relevant for the n-induced
rate, the K and K$^*$ amplitudes have the same sign, whereas in both
$T=0$ channels the interference between the two strange mesons
is destructive and, as a consequence,
the p-induced rate is lowered with respect to the n-induced rate.
Using the Nijmegen strong couplings constants leads to a final ratio that is
34\% smaller than the pion-only ratio, while using
the J\"ulich couplings leave this ratio unchanged,
due again mostly to their larger K$^*$ and $\omega$ couplings.
Employing the weak NNK couplings calculated with $\chi$PT
we obtain an increase of the  $\Gamma_n/\Gamma_p$ ratio by 17\% with
Nijmegen couplings while the ratio remains
unchanged for the J\"ulich model.

Even though it is not an observable, Table
\ref{tab:weakobs} also presents the ratio of PV to PC rates to aid in
the comparison with other theoretical calculations.
Again, adding the strange
mesons produces the largest effect, especially when using the J\"ulich
strong couplings.
The final PV/PC ratio is larger by more than a factor of two compared to
the pion-only ratio.
The results quoted in Ref. \cite{holstein} are of the order of 1
and, therefore, closer to our results obtained with the J\"ulich model.

The intrinsic asymmetry parameter, $a_\Lambda$, shown in Table
\ref{tab:weakobs} is also found to be very sensitive to the different
mesons included in the model.  This is the only observable which is
changed dramatically by the inclusion of the $\rho$, reducing the
pion-only value by more than a factor of two.  Adding the other mesons
increases $a_\Lambda$, leading to a result about 30\%
larger than for $\pi$-exchange alone in the case of the Nijmegen
couplings and 50\% larger for the J\"ulich model. The effect of using
the weak NNK couplings from $\chi$PT is very small for this
observable.

\subsection{Comparison with Experiment}

Our final results for various hypernuclei are presented in Table
\ref{tab:hyps}.  We find an overall agreement between our results
for the nonmesonic rate and the experimental values, especially when
the $\chi$PT weak couplings for the K meson are used, which yield
somewhat larger rates.

It has been the hope for many years that the inclusion of
additional mesons would dramatically increase the ratio of neutron- to
proton-induced rates. Here we find
the opposite to be true.
The final ratio greatly underestimates the newer
central experimental values, although the large experimental error
bars do not permit any definite conclusions at this time.
On the other hand, the proton-induced rate which has errors of the same
magnitude as the total rate is overpredicted by our calculations by up
to a factor of two.  It is the neutron-induced rate which has been very
difficult to measure accurately.  It is somewhat surprising that while
both individual rates appear in disagreement with the data their sum
conspires to a total rate which reproduces the measurements.
Other mechanisms
that have been explored to remedy this puzzle include quark-model
calculations which yield a large violation of the $\Delta I=1/2$
rule\cite{oka,maltman}, and the consideration of the 3N
emission
channel ($\Lambda N N \to N N N$) as a result of the pion being absorbed
on correlated 2N pairs\cite{albe,ramos1}.
A recent reanalysis\cite{ramos4},
which includes FSI of the three nucleons on their way out of the
nucleus via a Monte Carlo simulation, shows that the 2N-induced channel
further increases the experimental error bars and leads to
an experimental value
compatible with the predictions of the OPE model. However, the
same reference shows that a comparison of the calculated proton spectrum
with the experimental one favors values of $\Gamma_n/\Gamma_p=$2--3.
It is therefore
imperative, before speculating further about the deficiencies
of the present
models in reproducing this ratio, to carry out more precise
experiments
such as the measurement of the number of protons emitted per
$\Lambda$ decay, suggested in Ref. \cite{ramos4}.

Regarding the asymmetry parameter,
comparison with experiment can only be made at the level of the
measured proton asymmetry.
As discussed in Section II, this quantity is determined as a product of
the asymmetry parameter $A_p$, characteristic of the weak
decay, and the polarization of the hypernucleus, $P_y$, which must
be determined theoretically.
The energy resolution of the
experiment measuring the decay of polarized $^{12}_\Lambda$C
produced in a $(\pi^+,K^+)$ reaction\cite{ajim} was $5 - 7$ MeV which
did not allow
distinguishing between the first three $1^-$ states. Before the weak decay
occurs, the two excited states
decay electromagnetically to the ground state. Therefore, in order
to determine the polarization at this stage, one requires:
i) the polarization of the ground and
excited states, together with the corresponding formation cross
sections, and ii) an attenuation coefficient to account for the
loss of polarization in the transition of the excited states to the
ground state.  In Ref. \cite{iton2}, hypernuclear production cross
sections and polarizations have been estimated for the $(\pi^+,K^+)$
reaction in the distorted wave impulse approximation with
configuration-mixed wave functions. We note that the sum of the
cross sections for the two excited $1^-$ states amounts to 40\%
relative to
the ground state peak, which is consistent with the $(31\pm
8)$\% obtained in a fit to the Brookhaven $^{12}{\rm
C}(\pi^+,K^+)^{12}_\Lambda$C spectrum\cite{millener}. Using the values
of Ref. \cite{iton2} for the polarization and cross sections of
the $1^-$ states
in $^{12}_\Lambda$C together with the spin depolarization formalism
of Ref. \cite{ejiri}, we obtain
$P_y=-0.19$. This value,
together with
$A_p=0.151$ (Nijmegen) or 0.175 (J\"ulich),
determined from $a_\Lambda$ using Eq. (\ref{eq:asym}), leads to
an asymmetry ${\cal A}=-0.029$ (Nijmegen) or $-0.033$ (J\"ulich),
which lies within the uncertainties of the experimental
result.

Using the same model \cite{iton2} for
$^{11}_\Lambda$B, which predicts equal formation cross sections for
the $2^+_1$ and $2^+_2$ states, we obtain a polarization of
$P_y=-0.29$. However, hypernuclear structure calculations by
Auerbach {\it et al} \cite{auerbach} predicted strong configuration
mixing which
reduced the cross section of the lower $2^+$ state by a factor of
three relative to the higher one. This prediction was verified by a
reanalysis of older emulsion data\cite{dalitz2}. Taking these relative
weights into account, we obtain the value $P_y=-0.43$, which is the
one used in Table \ref{tab:hyps} and leads to better agreement with
the experimental asymmetry.
Just as in the case of the proton- to neutron-induced
ratio, the present level of uncertainty in the experiment
does not yet permit using the asymmetry as an observable that
differentiates between different models for the weak decay.

In order to avoid the need for theoretical input and access $A_p$
directly, a new experiment at KEK\cite{kishi} is measuring
the decay of polarized $^{5}_\Lambda$He, extracting both
the pion asymmetry from the mesonic
channel, ${\cal A}_{\pi^-}$, and the proton asymmetry from the
nonmesonic decay, ${\cal A}$.
The asymmetry parameter $a_{\pi^-}$ of the pionic channel has
been estimated to be very similar to that of the free $\Lambda$
decay\cite{motoba}, and, therefore, the hypernuclear polarization can now be
obtained from the relation $P_y={\cal A}_{\pi^-}/a_{\pi^-}$.
This in turn can then be used as input,
together with the measured value of ${\cal A}$, to determine the
asymmetry parameter for the nonmesonic decay from the equality
$A_p={\cal
A}/P_y$. This experiment will not only allow a clean extraction of
the nonmesonic asymmetry parameter but will also check theoretical model
predictions for the amount of hypernuclear polarization.

Finally, we briefly compare our results to previous calculations.
The only other shell model calculations we are aware of are
those of Ref. \cite{holstein} and Ref. \cite{kisslinger}. The
results of Ref.\cite
{kisslinger} are in agreement with ours while the preliminary results reported
by Dubach et al.\cite{holstein} appear to be very different.
Their uncorrelated OPE-only rate for $^{12}_\Lambda$C is listed as 3.4 which is
 about a
factor of 2 larger than ours while adding initial SRC, FF, and FSI reduce this
rate to 0.5.  This amounts to a reduction factor of almost 7, in contrast to
our suppression of roughly a factor of two. Furthermore, their correlated
rate for $^{5}_\Lambda$He is listed as 0.9, almost a factor of two larger than
 the $^{12}_\Lambda$C
result. Unfortunately, no details are given in Ref. \cite{holstein}
that address these problems.
We note, however, that there are some unexplained inconsistencies
between their recent results of Ref. \cite{holstein} and what was reported ten
years before in Ref. \cite{dubachwein}, where the correlated
$\pi$-exchange decay rate for $^{12}_\Lambda$C is 2.0 while the
addition of the
other mesons lowers this value to 1.2. In fact, these values are more
consistent with their $^5_\Lambda$He results, as well as with the
effect of short range correlations found in almost all studies of the
nonmesonic weak decay either in nuclear matter or finite nuclei.
Our results for $\Gamma_n/\Gamma_p$ again differ from what it is
reported in Ref.
\cite{holstein}, where a value
$\Gamma_n/\Gamma_p=0.2$ is obtained for $\pi$-exchange alone but
$0.83$ when all the mesons are included. However, their results
in finite nuclei are, surprisingly, quite different
from their nuclear matter results, namely
$\Gamma_n/\Gamma_p=0.06$ for $\pi$-exchange alone and 0.345 when
all mesons are included.
With regard to the asymmetry parameter, the nuclear
matter results of Ref.
\cite{holstein} are
qualitatively similar to our results for the Nijmegen couplings. They
obtain a value $a_\Lambda=-0.192$ for $\pi$-exchange alone and $-0.443$ when
all the mesons are considered.

\section{Conclusions}

In this study we have presented calculations for the weak nonmesonic decay
mode $\Lambda N \to NN$ of $\Lambda$-hypernuclei.
In contrast to most previous investigations performed in nuclear matter
this work analysed this hypernuclear decay in a nonrelativistic shell model
approach.  The initial hypernuclear and final nuclear structure are
taken into account through spectroscopic factors.
All possible initial and
final relative orbital angular momenta are included in the baryon-baryon
system. Realistic initial and final short-range correlations (SRC)
obtained from $\Lambda$N and NN interactions based on the
Nijmegen baryon-baryon potential are employed in order to treat
the nuclear structure details with
as few approximations and ambiguities as possible.
Our calculations were performed in a one-boson-exchange model
that includes not only the long-ranged pion but also contributions from the
other pseudoscalar mesons, the $\eta$ and K, as well as the vector
mesons $\rho, \omega$ and K$^*$.  The weak baryon-baryon-meson
vertices
were obtained using SU(6) and soft meson theorems for the PV vertices
and the pole model for the PC vertices.
The primary goal of this work was to reduce nuclear structure uncertainties
as much as possible so that our framework can be used to
extract these weak baryon-baryon-meson couplings.

Total decay rates evaluated with the full weak OBE potential fall within
15\% of the value obtained with pion exchange only and reproduce the
experimental data.  This is due to the interference between the
contributions of the heavier mesons whose individual influence on the
decay rate can be substantial. Including the kaon exchange alone reduces
the total rate by almost 50\%, this reduction is compensated by adding the
other mesons, specifically the K$^*$.  In contrast to previous studies we
found little influence from the $\rho$-meson even when we used a
different model for the weak $\Lambda {\rm N} \rho$ couplings,
similarly the
$\omega$ contributes at the 10\% level.
The dominant contribution
beyond the pion-exchange mechanism is clearly the kaon exchange,
followed by the K$^*$ which tends to partially cancel the effects of
the kaon.  It is therefore imperative that future studies include both
strange mesons simultaneously.  The importance of kaon exchange makes it
possible to see the effects of modifying the weak NNK couplings by
loop
contributions from next-to-leading order $\chi$PT.  Including these loop
graphs leads to a reduction of the NNK couplings from their tree-level
value up to 50\%, which in turns modifies the rates by up to
20\%. Future experiments should be able to verify this effect.

We found the dominance of strange mesons to be even more pronounced in
the partial rates and their ratio.  Including the kaon
reduces this ratio by more than a factor of three, which again is
compensated by the K$^*$.  Furthermore, this ratio turns out to be
sensitive to the choice of strong coupling constants as well. Using the
Nijmegen strong couplings reduces the ratio by 30\% from its pion-only
value while the use of the J\"ulich strong couplings leads to a change
of only a few percent.  This finding indicates the need for improved YN
potentials with better determined strong couplings at the
hyperon-nucleon-meson vertices.  Both theoretical
values are far away from the experimental
data, even though the error bars are still large.  It appears to be
impossible to reconcile these discrepancies within a one-boson-exchange
potential.  In order to approach the experimental values the weak
couplings of the heavier mesons would have to be unreasonably large
which would yield very large total rates incompatible with the data.
If future experiments with improved partial rates confirm the present
trend new mechanisms of a different kind would
have to be introduced to resolve this puzzle.
In contrast to the previous observables we found the proton asymmetry to
be very sensitive to the $\rho$-exchange while the influence of the
kaon is more moderate.  This polarization observable is therefore an
important addition to the set of observables since its sensitivities are
different from the total and partial rates.  The results of our
calculations are within the very large current error bounds.

\subsection{Criticisms}

Our study clearly indicates that further theoretical
effort must be invested to understand the dynamics of the nonmesonic
weak hypernuclear decay.
Within the one-meson exchange picture it would be desirable to use weak
coupling constants developed with more sophisticated approaches.  A
beginning has been made by Savage and Springer\cite{spring}
in their evaluation of
the weak NNK couplings in next-to-leading order $\chi$PT and the
effect
has been found to be important.  However, an understanding of the weak
$\Lambda {\rm N} \pi$ and $\Sigma {\rm N} \pi$ couplings within the
framework
of chiral lagrangians is still missing.  Furthermore, due to the
importance of the K$^*$-meson it would be desirable to recalculate its
weak NNK$^*$ couplings in improved models as well.
Several recent studies\cite{oka} have gone beyond the
conventional picture of meson exchange
and have developed mechanisms based purely on
quark degrees of freedom.  One should keep in mind, however,
that such models have not always been able to reproduce the experimentally
measured free hyperon decays.

Another avenue that is currently being pursued is the validity of the
$\Delta$I=1/2 rule in the $\Lambda N \to NN$ process.  While this empirical rule is
well established for the free hyperon and kaon decays there is some
indication that it could be violated for the $\Lambda N \to NN$ process.
Within the
framework of SU(3) and soft meson theorems the weak vertices of NNK and
$\Lambda {\rm N} \eta$ are related to the observable $\Lambda {\rm N}
\pi$ decay,
therefore, one would expect small $\Delta I$=3/2 contributions for these
mesons. On the other hand, the vector meson vertices can receive
substantial contributions from factorization terms which have been shown
not to fulfil the $\Delta$I=1/2 rule\cite{maltman}.  The attractive
feature of these additional terms is their strong influence on
the ratio $\Gamma_n/\Gamma_p$.

On the level of implanting the basic $\Lambda N \to NN$ amplitude into the
 nucleus
uncertainties have been minimized our study by treating each ingredient
as well as possible. Nevertheless, within the framework of the impulse
approximation and the shell model the short range correlations, the
spectroscopic factors and the single particle wavefunctions still come
from separate sources.  This dilemma can be avoided in rigorous few-body
calculations with realistic wave functions.  The nonmesonic decay of the
hypertriton can be calculated using correlated three-body hypernuclear
wave functions for the initial hypertriton state and continuum Faddeev
solutions for the three-nucleon scattering state\cite{golak}.
Thus, all nuclear structure input is generated from the same underlying
YN- and NN-potentials, eliminating the ambiguities of the shell model
approach.  It is therefore of utmost importance to pursue experimental
measurements of the nonmesonic decay of the hypertriton.

\subsection{Outlook}

On the experimental side,
it is critical to obtain new high
accuracy data soon.  Improved partial rates for the proton- and
neutron-induced decay modes are especially important. Of help would be
to not only measure rates but also exclusive spectra of the decay
products. Such distributions would be significant to disentagle the
effects of the $\Lambda N N\to NNN$ process from the two-body process discussed
 here.
Beyond improving the present data base for the weak decay of
$\Lambda$-hypernuclei, there are two more avenues which would aid our
understanding of the weak $\Delta S$=1 hadronic interaction.

First, with the advent of new, high precision proton accelerators such
as COSY in J\"ulich, it may become possible to perform a direct study of
the time-reversed process $p n \to \Lambda p$ \cite{haidenbauer}.  While
the very low cross sections present in this direct investigation of the
$\Delta S$=1 baryon-baryon interaction will be difficult to measure,
high efficiency
detection schemes should allow determining a branching ratio of
$10^{-13}$.  Thus, the strangeness changing hadronic weak interaction
could be studied similarly to the weak parity-violating NN interaction.
The asymmetry of this reaction has been measured at several kinematics
which are sensitive to different parts of the meson exchange potential.
Furthermore, measuring the $p n \to \Lambda p$ process directly would
give access to a number of polarization observables since the $\Lambda$
is self-analyzing.

Secondly, the hypernuclear weak decay studies should be extended to
double-$\Lambda$ hypernuclei.  Very few events involving
these exotic objects --- whose
very existence would place stringent constraints on the existence of the
elusive H-dibaryon --- have been reported.  Studying the weak decay of
these objects would open the door to a number of new exotic
$\Lambda$-induced decays: $\Lambda \Lambda \to \Lambda N$ and $\Lambda
\Lambda \to \Sigma N$.  Both of these decays would involve hyperons in
the final state and should be distinguishable from the ordinary $\Lambda N \to
 NN$
mode.  Especially the $\Lambda \Lambda \to \Lambda N$ channel would be
intruiging since the dominant pion exchange is forbidden, thus this
reaction would have to occur mostly through kaon exchange.  One would
therefore gain access to the $\Lambda \Lambda$K vertex.

Even with the demise of KAON, the promising efforts at KEK
with an improved measurement of the $^{5}_\Lambda$He decay,
the continuing program at BNL,
and the advent of
the hypernuclear physics program (FINUDA) at DA$\Phi$NE represent
excellent opportunities to obtain new valuable information
that will shed light onto the
still unresolved problems of the weak decay of hypernuclei.

\acknowledgements
We are grateful to Prof. P. Pascual for clarifying discussions regarding
sign conventions and relations between coupling constants.
The work of CB was supported by US-DOE grant no. DE-FG02-95-ER40907 while
the work of AP and AR was supported by
DGICYT contract no. PB92-0761 (Spain) and by the Generalitat de
Catalunya grant no. GRQ94-1022.
This work has received support from the NATO Grant
CRG 960132.
AP acknowledges support
from a doctoral fellowship of the Ministerio de Educaci\'on y Ciencia
(Spain).

\pagebreak

\section{Appendix}
\subsection{Coefficients $\langle (L' S) J M_J | {\hat O}_\alpha
| (L_r S_0) J M_J \rangle$}
\begin{itemize}
\item{Spin-Spin transition}
\begin{equation}
\langle (L' S) J M_J | {\hat O}_\alpha | (L_r S_0) J M_J \rangle
= (2S(S+1)-3) \,\,\, \delta_{L_r L'}
\,\,\, \delta_{S_i S}
\label{coefcent}
\end{equation}
\item{Tensor transition}
\begin{equation}
\langle (L' S) J M_J | {\hat O}_\alpha | (L_r S_0) J M_J \rangle
= S^{J}_{L_r L'} \,\,\, \delta_{S_0 S}
\,\,\, \delta_{S 1} \ ,
\label{coeftens}
\end{equation}
where the coefficients $S^{J}_{L_r L'}$ are given in Table
\ref{tab:coeften}.
\item{PV transition}
\\
\\
{\em Pseudoscalar mesons}
\begin{eqnarray}
\langle (L' S) J M_J | {\hat O}_\alpha | (L_r S_0) J M_J \rangle
&=& (-1)^{J+1-L'} \sqrt{6} \sqrt{2S_0+1}
\sqrt{2L_r+1} \sqrt{2S+1} \nonumber \\
& & \langle 1 0 L_r 0 | L' 0 \rangle\,\,\,\,\,\left(
\begin{array}{ccc}
\frac{1}{2} & \frac{1}{2} & S_0 \\
S & 1 & \frac{1}{2}
\end{array}
\right)
\left(
\begin{array}{ccc}
L' & L_r & 1 \\
S_0 & S & J
\end{array}
\right) \nonumber \\
\label{coefpvps}
\end{eqnarray}
{\em Vector mesons}
\begin{eqnarray}
\langle (L' S) J M_J | {\hat O}_\alpha | (L_r S_0) J M_J \rangle
&=& i (-1)^{J-L'+S} 6 \sqrt{6} \sqrt{2S_0+1}
\sqrt{2L_r+1} \sqrt{2S+1} \nonumber \\
& & \langle 1 0 L_r 0 | L' 0 \rangle\,\,\,\,\,
\left(
\begin{array}{ccc}
L' & L_r & 1 \\
S_0 & S & J
\end{array}
\right)
\left(
\begin{array}{ccc}
1 & 1 & 1  \\
\frac{1}{2} & \frac{1}{2} & S \\
\frac{1}{2} & \frac{1}{2} & S_0
\end{array}
\right) \nonumber \\
\label{coefpvvec}
\end{eqnarray}
\end{itemize}

\begin{table}
\centering
\caption{Possible $^{2S+1}L_J$ channels obtained in the weak decay
of $p$-shell hypernuclei}
\vspace{0.5cm}
\begin{tabular}{lcccc}
 $L_r$ & Weak decay channel & $L'$ &Strong FSI& $L$ \\
        & ($\Lambda$N) & (NN) & (NN) \\
\hline
    & Central  &   &  &\\
\hline
 $^1S_0$ & $\to$    & $^1S_0$ & $\to$ & $^1S_0$ \\
 $^3S_1$ & $\to$    & $^3S_1$ & $\to$ & $^3S_1$, $^3D_1$ \\
 $^1P_1$ & $\to$    & $^1P_1$ & $\to$ & $^1P_1$ \\
 $^3P_0$ & $\to$    & $^3P_0$ & $\to$ & $^3P_0$ \\
 $^3P_1$ & $\to$    & $^3P_1$ & $\to$ & $^3P_1$ \\
 $^3P_2$ & $\to$    & $^3P_2$ & $\to$ & $^3P_2$, $^3F_2$ \\
\hline
    & Tensor  &   &  &\\
\hline
 $^3S_1$ & $\to$    & $^3D_1$ & $\to$ & $^3D_1$, $^3S_1$ \\
 $^3P_0$ & $\to$    & $^3P_0$ & $\to$ & $^3P_0$ \\
 $^3P_1$ & $\to$    & $^3P_1$ & $\to$ & $^3P_1$ \\
 $^3P_2$ & $\to$    & $^3P_2$, $^3F_2$ & $\to$ & $^3P_2$, $^3F_2$ \\
\hline
    & P.V.  &   &  &\\
\hline
 $^1S_0$ & $\to$   & $^3P_0$ & $\to$ & $^3P_0$ \\
 $^3S_1$ & $\to$   & $^1P_1$ & $\to$ & $^1P_1$ \\
 $^3S_1$ & $\to$   & $^3P_1$ & $\to$ & $^3P_1$ \\
 $^1P_1$ & $\to$   & $^3S_1$, $^3D_1$ & $\to$ & $^3S_1$, $^3D_1$ \\
 $^3P_0$ & $\to$   & $^1S_0$ & $\to$ & $^1S_0$ \\
 $^3P_1$ & $\to$   & $^3S_1$, $^3D_1$ & $\to$ & $^3S_1$, $^3D_1$ \\
 $^3P_2$ & $\to$   & $^1D_2$ & $\to$ & $^1D_2$ \\
 $^3P_2$ & $\to$   & $^3D_2$ & $\to$ & $^3D_2$
\end{tabular}
\label{tab:chan}
\end{table}

\begin{table}
\centering
\caption{Constants appearing in weak transition potential for the
different mesons (in units of $G_F {m_\pi}^2 = 2.21 \times 10^{-7} $)}
\vskip 0.2 in
\begin{tabular}{lcccc}
 $\mu_i$  & $K^{(i)}_{C}$ & $K^{(i)}_{SS}$ & $K^{(i)}_{T}$ &
$K^{(i)}_{PV}$
\\
\hline\hline
&   &    &       &         \\
$\pi$ & 0 & $\displaystyle\frac{B_\pi}{2 \overline M}
\displaystyle\frac{g_{\rm {\scriptscriptstyle NN} \pi}}{2M}$ &
 $\displaystyle\frac{B_\pi}{2 \overline M}
\displaystyle\frac{g_{\rm {\scriptscriptstyle NN} \pi}}{2M}$ &
$A_\pi  \displaystyle\frac{g_{\rm {\scriptscriptstyle NN} \pi}}{2M}$ \\
&   &    &       &         \\
$\eta$ & $0$ & $\displaystyle\frac{B_\eta}{2 \overline M}
\displaystyle\frac{g_{\rm {\scriptscriptstyle NN} \eta}}{2M}$ &
$\displaystyle\frac{B_\eta}{2 \overline M}
\displaystyle\frac{g_{\rm {\scriptscriptstyle NN} \eta}}{2M}$ &
$A_\eta  \displaystyle\frac{g_{\rm {\scriptscriptstyle NN} \eta}}{2M}$ \\
&    &     &       &         \\
K & $0$ & $\displaystyle\frac{1}{2M} \displaystyle\frac{g_{\rm
 {\scriptscriptstyle \Lambda N K}}}
{2 \overline M}$ &
$\displaystyle\frac{1}{2 M}
\displaystyle\frac{g_{\rm \scriptscriptstyle{\Lambda N K}}}{2 \overline M}$
& $\displaystyle\frac{g_{\rm \scriptscriptstyle{\Lambda N K}}}{2 M}$
\\
&    &     &       &         \\
$\rho$ & $g^{\rm {\scriptscriptstyle V}}_{\rm {\scriptscriptstyle NN} \rho}
 \alpha_\rho$ &
$2\displaystyle\frac{\alpha_\rho + \beta_\rho}{2 \overline M}
\displaystyle\frac{g^{\rm {\scriptscriptstyle V}}_{\rm {\scriptscriptstyle NN}
 \rho} +
g^{\rm {\scriptscriptstyle T}}_{\rm {\scriptscriptstyle NN} \rho}} {2M}$ &
$ - \displaystyle\frac{\alpha_\rho + \beta_\rho}{2 \overline M}
\displaystyle\frac{ g^{\rm {\scriptscriptstyle V}}_{\rm {\scriptscriptstyle NN}
 \rho} +
g^{\rm {\scriptscriptstyle T}}_{\rm {\scriptscriptstyle NN} \rho}} {2M}$ &
$ - \varepsilon_\rho
\displaystyle\frac{ g^{\rm {\scriptscriptstyle V}}_{\rm {\scriptscriptstyle NN}
 \rho} +
g^{\rm {\scriptscriptstyle T}}_{\rm {\scriptscriptstyle NN} \rho}} {2M}$ \\
&    &     &       &         \\
$\omega$ & $g^{\rm {\scriptscriptstyle V}}_{\rm {\scriptscriptstyle NN} \omega}
 \alpha_\omega$ &
$2\displaystyle\frac{\alpha_\omega + \beta_\omega}{2 \overline M}
\displaystyle\frac{g^{\rm {\scriptscriptstyle V}}_{\rm {\scriptscriptstyle NN}
 \omega} +
g^{\rm {\scriptscriptstyle T}}_{\rm {\scriptscriptstyle NN} \omega}} {2M} $ &
$ - \displaystyle\frac{\alpha_\omega + \beta_\omega}{2 \overline M}$
$\displaystyle\frac{ g^{\rm {\scriptscriptstyle V}}_{\rm {\scriptscriptstyle NN}
 \omega} +
g^{\rm {\scriptscriptstyle T}}_{\rm {\scriptscriptstyle NN} \omega}} {2M} $ &
$ - \varepsilon_\omega
\displaystyle\frac{ g^{\rm {\scriptscriptstyle V}}_{\rm {\scriptscriptstyle NN}
 \omega} +
g^{\rm {\scriptscriptstyle T}}_{\rm {\scriptscriptstyle NN} \omega}} {2M}$ \\
&    &     &       &         \\
K$^*$ & $g^{\rm {\scriptscriptstyle V}}_{\rm {\scriptscriptstyle
\Lambda N K^*}} $ &
$2 \displaystyle\frac{1}{2M} \displaystyle\frac{g^{\rm
 \scriptscriptstyle{V}}_{\rm {\scriptscriptstyle \Lambda N K^*}} +
g^{\rm \scriptscriptstyle{T}}_{\rm {\scriptscriptstyle \Lambda N K^*}}} {2
 \overline M}$  &
$ - \displaystyle\frac{1}{2 M}
\displaystyle\frac{ g^{\rm {\scriptscriptstyle V}}_{\rm {\scriptscriptstyle
 \Lambda N K^*}} +
g^{\rm \scriptscriptstyle{T}}_{\rm {\scriptscriptstyle \Lambda N K^*}}} {2
 \overline M}$ &
$- \displaystyle\frac{ g^{\rm \scriptscriptstyle{V}}_{\rm {\scriptscriptstyle
 \Lambda N K^*}} +
g^{\rm \scriptscriptstyle{T}}_{\rm {\scriptscriptstyle \Lambda N K^*}}} {2 M}$
 \\
&    &     &       &         \\
\end{tabular}
\label{tab:k}
\end{table}

\begin{table}
\centering
\caption{Nijmegen (J\"ulich) strong and weak coupling constants
and cutoff parameters for the different mesons.
The weak couplings are in units of $G_F {m_\pi}^2 = 2.21 \times 10^{-7} $.
For the kaon and the
$\rho$-meson we also quote the weak couplings obtained by
Ref. \protect\cite{spring} and Ref. \protect\cite{nardulli},
respectively.}
\vskip 0.2 in
\begin{tabular}{|c|l|l|l|l|}
Meson & \multicolumn{1}{c|}{Strong c.c.} & \multicolumn{2}{c|}{Weak c.c.} &
$\Lambda_i$ \\
      &     & \multicolumn{1}{c}{PC} & \multicolumn{1}{c|}{PV} &
\mbox{(GeV)} \\ \hline
 $\pi$ & $g_{\rm {\scriptscriptstyle NN} \pi}$ = 13.3 & $B_\pi$=$-$7.15 &
$A_\pi$=1.05 & 1.30  \\
    & $g_{\rm {\scriptscriptstyle \Lambda \Sigma} \pi}$ = 12.0 &  &  &  \\
\hline
 $\eta$ & $g_{\rm {\scriptscriptstyle NN} \eta}$ = 6.40(0.) & $B_\eta$=$-$14.3 &
 $A_\eta$=1.80 & 1.30 \\
   & $g_{\rm {\scriptscriptstyle \Lambda \Lambda} \eta}= -6.56(0.)$ &  &  &  \\
\hline
K & $g_{\rm {\scriptscriptstyle \Lambda NK}}$ = $-$14.1($-$13.5) &
$C_{\rm {\scriptscriptstyle K}}^{\rm {\scriptscriptstyle PC}}$=$-$18.9
& $C_{\rm {\scriptscriptstyle K}}^{\rm {\scriptscriptstyle PV}}$=0.76 & 1.20 \\
   &   &\phantom{$C_{\rm {\scriptscriptstyle K}}^{\rm {\scriptscriptstyle
 PC}}$}=$-14.0$
\cite{spring} &
\phantom{$C_{\rm {\scriptscriptstyle K}}^{\rm {\scriptscriptstyle PV}}$}=$0.40$
\cite{spring} & \\
    & $g_{\rm {\scriptscriptstyle N \Sigma K}}$ = 4.28(3.55) & $D_{\rm
 {\scriptscriptstyle K}}^
{\rm {\scriptscriptstyle PC}}$=6.63 & $D_{\rm {\scriptscriptstyle K}}^{\rm
 {\scriptscriptstyle PV}}$=2.09  & \\
   &   &\phantom{$D_{\rm {\scriptscriptstyle K}}^{\rm {\scriptscriptstyle
 PC}}$}=3.20
\cite{spring} &
\phantom{$D_{\rm {\scriptscriptstyle K}}^{\rm {\scriptscriptstyle PV}}$}=1.50
\cite{spring} &  \\
\hline
$\rho$ & $g_{\rm {\scriptscriptstyle NN} \rho}^{\rm {\scriptscriptstyle V}}$ =
 3.16(3.25) &
$\alpha_\rho$=$-$3.50 & $\epsilon_\rho$=1.09  & 1.40 \\
   &  & \phantom{$\alpha_\rho$}=$-3.39$\cite{nardulli} &
\phantom{$\epsilon_\rho$}=3.84\cite{nardulli} & \\
       & $g_{\rm {\scriptscriptstyle NN} \rho}^{\rm {\scriptscriptstyle T}}$ =
 13.3(19.8) &
$\beta_\rho$=$-$6.11   &  &  \\
   &  & \phantom{$\beta_\rho$}=$-7.11$\cite{nardulli} &  &  \\
   & $g_{\rm {\scriptscriptstyle \Lambda \Sigma} \rho}^{\rm {\scriptscriptstyle
 V}}= 0(0)$ &
   &  & \\
   & $g_{\rm {\scriptscriptstyle \Lambda \Sigma} \rho}^{\rm {\scriptscriptstyle
 T}}=
   11.2(16.0)$ &  &  &  \\
\hline
$\omega$ & $g_{\rm {\scriptscriptstyle NN} \omega}^{\rm {\scriptscriptstyle V}}$
 = 10.5(15.9)
& $\alpha_\omega$=$-$3.69  & $\epsilon_\omega$= $-$1.33
& 1.50 \\
      & $g_{\rm {\scriptscriptstyle NN} \omega}^{\rm {\scriptscriptstyle T}}$ =
 3.22(0) &
$\beta_\omega$=$-$8.04  &  & \\
   & $g_{\rm {\scriptscriptstyle \Lambda \Lambda} \omega}^{\rm
 {\scriptscriptstyle V}}=
   7.11(10.6)$ & &  &  \\
   & $g_{\rm {\scriptscriptstyle \Lambda \Lambda} \omega}^{\rm
 {\scriptscriptstyle T}}=
   -4.04(-9.91)$ & &  &  \\
\hline
K$^*$ & $g_{\rm {\scriptscriptstyle \Lambda N K^*}}^{\rm
{\scriptscriptstyle V}}$ =
$-$5.47($-$5.63) & $C^{\rm {\scriptscriptstyle PC,V}}_{\rm {\scriptscriptstyle
K^*}}$=$-$3.61 &
$C^{\rm {\scriptscriptstyle PV}}_{\rm {\scriptscriptstyle K^*}}$=$-$4.48 & 2.20
 \\
     & $g_{\rm {\scriptscriptstyle \Lambda N K^*}}^{\rm {\scriptscriptstyle T}}$
 =
$-$11.9($-$18.4) &
$C^{\rm {\scriptscriptstyle PC,T}}_{\rm {\scriptscriptstyle K^*}}$=$-$17.9
&   & \\
   &$g_{\rm {\scriptscriptstyle N \Sigma K^*}}^{\rm {\scriptscriptstyle
 V}}=-3.16(-3.25)$ &
   $D^{\rm {\scriptscriptstyle PC,V}}_{\rm {\scriptscriptstyle K^*}}$=$-$4.89 &
   $D^{\rm {\scriptscriptstyle PV}}_{\rm {\scriptscriptstyle K^*}}$=0.60 & \\
   &$g_{\rm {\scriptscriptstyle N \Sigma K^*}}^{\rm {\scriptscriptstyle
 T}}=6.00(7.87)$ &
   $D^{\rm {\scriptscriptstyle PC,T}}_{\rm {\scriptscriptstyle K^*}}$=9.30 &  &
 \\
\end{tabular}
\label{tab:const}
\end{table}

\begin{table}
\centering
\caption{$\pi$ exchange contribution to the
$\Lambda N \to N N$
decay rate of ${}^{12}_\Lambda$C }
\vskip 0.1 in
\begin{tabular}{lcccccc}
   & FREE & SRC & SRC+FF & \multicolumn{3}{c}{SRC+FF+FSI} \\
   &   &   &  & phenom. & Nijm93 & Reid93 \\
   &   &   &  & Eq. (\ref{eq:cor}) &   &   \\
\hline\hline
C  & 0.282 & $3.4 \times 10^{-3}$ & $1.3 \times 10^{-2}$ &
$4.2 \times 10^{-3}$ & $3.3 \times 10^{-3}$ & $4.0 \times 10^{-3}$ \\
T  & 0.858 & 0.781 & 0.637 & 0.685 & 0.566 & 0.579 \\
PC & 1.140 & 0.785 & 0.650 & 0.689 & 0.531 & 0.547 \\
PV & 0.542 & 0.447 & 0.389 & 0.421 & 0.353 & 0.345 \\
\hline
$\Gamma/\Gamma_\Lambda$ & 1.682 & 1.232 & 1.038 & 1.110 & 0.885 &
0.892 \\
\hline\hline
$\Gamma_n/\Gamma_p$ & 0.182 & 0.113 & 0.120 & 0.118 & 0.104 &
0.100 \\
\hline
PV/PC & 0.476 & 0.570 & 0.598 & 0.610 & 0.665 & 0.631 \\
\hline
$a_\Lambda$ & $-0.594$ & $-0.420$ & $-0.506$ & $-0.484$ &
$-0.238$ & $-0.242$ \\
\end{tabular}
\label{tab:pi}
\end{table}

\begin{table}
\centering
\caption{$\pi$ and $\rho$ exchange contribution to the
$\Lambda N \to N N$
decay rate of ${}^{12}_\Lambda$C.
The values in parentheses have been calculated using
the weak coupling constants by Nardulli\protect\cite{nardulli}.}
\vskip 0.1 in
\begin{tabular}{lccc}
\ & $\pi$ & $\rho$ & $\pi+\rho$ \\
\hline\hline
C (C) & --- & 0.020 (0.019) & 0.020 (0.019) \\
C (SS) & 0.003 & 0.014 (0.016) & 0.006 (0.007) \\
C (Total)  & 0.003 & 0.045 (0.047) & 0.028 (0.030) \\
T        & 0.566 & 0.027 (0.032) & 0.373 (0.358) \\
PC       & 0.531 & 0.021 (0.023) & 0.445 (0.428) \\
PV       & 0.353 & 0.008 (0.096) & 0.414 (0.635) \\
\hline
$\Gamma/\Gamma_\Lambda$ & 0.885 & 0.029 (0.120) & 0.859 (1.063) \\
\hline
$\Gamma_n/\Gamma_p$ & 0.104 & 0.076 (0.097) & 0.095 (0.063) \\
\hline
$a_\Lambda$ &-0.238 & $ 0.036$ ($ 0.046$)& $-0.100$ ($-0.008$)  \\
\end{tabular}
\label{tab:pirho}
\end{table}

\begin{table}
\centering
\caption{K and K$^*$ exchange contribution to the
$\Lambda N \to N N$
decay rate of ${}^{12}_\Lambda$C.
The values in parentheses have been calculated using the
the NNK weak coupling constants obtained from next-to-leading order in
$\chi$PT\protect\cite{spring}.}
\vskip 0.1 in
\begin{tabular}{lccc}
\ & K & K$^*$ & ${\rm K}+{\rm K}^*$ \\
\hline\hline
C (C) & --- & 0.019 & 0.019 (0.019) \\
C (SS) & 0.004 (0.002) & 0.092 & 0.130 (0.122) \\
C (Total)  & 0.004 (0.002) & 0.038 & 0.063 (0.058) \\
T        & 0.083 (0.038) & 0.038 & 0.015 (0.005) \\
PC       & 0.093 (0.044) & 0.037 & 0.082 (0.050) \\
PV       & 0.040 (0.018) & 0.023 & 0.091 (0.061) \\
\hline
$\Gamma/\Gamma_\Lambda$ & 0.133 (0.062) & 0.060 & 0.173 (0.111) \\
\hline
$\Gamma_n/\Gamma_p$ & 0.263 (0.272) & 0.500 & 0.647 (0.760) \\
\hline
$a_\Lambda$ & $-0.080$ ($-0.090$) & $-0.192$ & $-0.426$ ($-0.532$)  \\
\end{tabular}
\label{tab:kakst}
\end{table}

\begin{table}
\centering
\caption{$\eta$ and $\omega$ exchange contribution to the
$\Lambda N \to N N$
decay rate of ${}^{12}_\Lambda$C.}
\vskip 0.1 in
\begin{tabular}{lccc}
\ & $\eta$ & $\omega$ & $\eta+\omega$ \\
\hline\hline
C (C) & --- & 0.045 & 0.045 \\
C (SS) & 0.001 & 0.009 & 0.016 \\
C (Total)  & 0.001 & 0.036 & 0.036 \\
T        & 0.005 & 0.004 & $1.5 \times 10^{-4}$ \\
PC       & 0.006 & 0.024 & 0.035 \\
PV       & 0.003 & 0.002 & 0.005 \\
\hline
$\Gamma/\Gamma_\Lambda$ & 0.009 & 0.026 & 0.041 \\
\hline
$\Gamma_n/\Gamma_p$ & 0.383 & 0.235 & 0.183 \\
\hline
$a_\Lambda$ & $-0.114$ & $-0.086$ & $-0.134$ \\
\end{tabular}
\label{tab:etaome}
\end{table}

\begin{table}
\centering
\caption{Free and fully correlated nonmesonic decay rate of
 ${}^{12}_\Lambda$C in units of the free $\Lambda$ decay rate
$\Gamma_\Lambda$. The values
in parentheses have been calculated using the J\"ulich-B coupling
constants at the strong vertex.}
\vskip 0.1 in
\begin{tabular}{lcc}
 & FREE & SRC+FF+FSI \\
\hline\hline
$\pi$ & 1.682 (1.682) & 0.885 (0.885) \\
$+ \rho$ & 2.055 (2.325) & 0.859 (0.831) \\
$+$K & 1.336 (1.699) & 0.497 (0.506) \\
$+$K$^*$ & 2.836 (3.821) & 0.760 (0.902) \\
$+ \eta$ & 2.467 (3.821) & 0.683 (0.902) \\
$+ \omega$ & 2.301 (4.338) & 0.753 (1.023) \\
\hline
 & & \\
\parbox{3cm}{weak K-couplings from $\chi$PT \cite{spring}} &
 & 0.844 (1.104) \\
 & & \\
\end{tabular}
\label{tab:rate}
\end{table}

\begin{table}
\centering
\caption{Weak decay observables
 for ${}^{12}_\Lambda$C. The values in parentheses
 have been calculated using the J\"ulich-B coupling
 constants at the strong vertex.}
\vskip 0.1 in
\begin{tabular}{lccc}
 & $\Gamma_n/\Gamma_p$ & PV/PC & $a_\Lambda$ \\
$\pi$ & 0.104 (0.104) & 0.665 (0.665) & $-0.238$ $(-0.238)$ \\
$+ \rho$ & 0.095 (0.096) & 0.930 (1.137) & $-0.100$ $(-0.052)$ \\
$+$K & 0.030 (0.029) & 2.413 (3.206) & $-0.138$ $(-0.074)$ \\
$+$K$^*$ & 0.049 (0.070) & 1.797 (1.968) & $-0.182$ $(-0.202)$ \\
$+ \eta$ & 0.058 (0.070) & 2.249 (1.968) & $-0.200$ $(-0.202)$ \\
$+ \omega$ & 0.068 (0.109) & 2.077 (1.675) & $-0.316$ $(-0.368)$ \\
\hline
 & & & \\
\parbox{3cm}{weak NNK-couplings from $\chi$PT \cite{spring}}
 & 0.080 (0.108) &
1.678 (1.436) & $-0.302$ $(-0.350)$ \\
 & & & \\
\end{tabular}
\label{tab:weakobs}
\end{table}
\begin{table}
\centering
\caption{Weak decay observables for various hypernuclei.
The values in parentheses have been calculated using the
NNK weak coupling constants obtained from next-to-leading order in
$\chi$PT\protect\cite{spring}.}
\vskip 0.1 in
\begin{tabular}{lccc}
 & $^5_\Lambda$He
 & $^{11}_\Lambda$B
 & $^{12}_\Lambda$C \\
\hline
$\Gamma/\Gamma_\Lambda$ & 0.414 (0.467) & 0.611 (0.686) & 0.753
(0.844) \\
EXP: & $0.41\pm 0.14$ \cite{szymanski} & $0.95 \pm 0.13 \pm 0.04$ \cite{noumi}
 & $1.14\pm 0.2$ \cite{szymanski} \\
 & & & $0.89 \pm 0.15 \pm 0.03$
\cite{noumi} \\
\hline
$\Gamma_n/\Gamma_p$ & 0.073 (0.089) & 0.084 (0.099) & 0.068 (0.080) \\
EXP: & $0.93\pm 0.55$\cite{szymanski} &
       $1.04^{+0.59}_{-0.48}$ \cite{szymanski} & $1.33^{1.12}_{-0.81}$
       \cite{szymanski} \\
     &  & $2.16 \pm 0.58^{+0.45}_{-0.95}$ \cite{noumi} &
          $1.87 \pm 0.59^{+0.32}_{-1.00}$ \cite{noumi} \\
     &  & $0.70\pm 0.3$\cite{montwill} & $0.70\pm 0.3$\cite{montwill} \\
     &  & $0.52\pm 0.16$\cite{montwill} & $0.52\pm 0.16$\cite{montwill}\\
\hline
$\Gamma_p/\Gamma_\Lambda$ & 0.386 (0.428) & 0.563 (0.624) & 0.705 (0.782) \\
EXP: & $0.21 \pm 0.07$\cite{szymanski} & $0.30^{+0.15}_{-0.11}$\cite{noumi} &
$0.31^{+0.18}_{-0.11}$\cite{noumi} \\
\hline
$a_\Lambda$ & $-0.273$ ($-0.264$) & $-0.391$ ($-0.378$) &
              $-0.316$ ($-0.302$) \\
\hline
${\cal A}(0^{\rm o})$ &  &$-0.120$ ($-0.116$) & $-0.030$ ($-0.029$) \\
EXP: & & $-0.20\pm 0.10$ \cite{ajim} & $-0.01\pm 0.10$ \cite{ajim} \\
\end{tabular}
\label{tab:hyps}
\end{table}

\begin{table}
\centering
\caption{Matrix elements of the tensor operator evaluated
between generalized spherical harmonic states of definite
J,L and S}
\vspace{0.5cm}
\begin{tabular}{lccc}
    &   &   &  \\
$S^{J}_{L_r L'}$ & $L'=J+1$ & $L'=J$ & $L'= J-1$ \\
    &   &   &  \\
\hline
    &   &   &  \\
$L_r=J+1$ & $\frac{-2(J+2)}{2J+1}$ & 0 & $\frac{6 \sqrt{J(J+1)}}
{2J+1}$ \\
    &   &   &  \\
$L_r=J$ & 0 & 2 & 0 \\
    &   &   &  \\
$L_r=J-1$ & $\frac{6 \sqrt{J(J+1)}}{2J+1}$ & 0 &
$\frac{-2(J-1)}{2J+1}$ \\
    &   &   &  \\
\end{tabular}
\label{tab:coeften}
\end{table}
\end{document}